\documentclass[11pt,a4paper]{article}
\pdfoutput=1 

\usepackage{jheppub} 

\usepackage[T1]{fontenc}
\usepackage[dvipsnames]{xcolor} 
\usepackage{caption}
\usepackage{booktabs}
\usepackage{subdepth}
\usepackage{dsfont}
\usepackage{amsmath}
\usepackage{subfig} 
\usepackage{makecell}
\usepackage{comment}
\usepackage{bm}
\usepackage{xstring}
\usepackage{import}

\newcommand{\SLJ}[3]{\ensuremath{{\:\!}^{{#1}\!}{#2}_{#3}}}
\newcommand{\chisq}[0]{\ensuremath{\chi^2/N_{\sf dof}}}
\newcommand{\grey}[1]{\textcolor{gray}{#1}}

\newcommand{\etaoct}[0]{\ensuremath{\eta_{\mathbf{8}}}}
\newcommand{\etas}[0]{\ensuremath{\eta_{\mathbf{1}}}}
\newcommand{\omoct}[0]{\ensuremath{\omega_{\mathbf{8}}}}
\newcommand{\oms}[0]{\ensuremath{\omega_{\mathbf{1}}}}
\newcommand{\fzs}[0]{\ensuremath{f_{0, \mathbf{1}}}}
\newcommand{\D}[0]{\ensuremath{D_{\bar{\mathbf{3}}}}}
\newcommand{\Dst}[0]{\ensuremath{D^*_{\bar{\mathbf{3}}}}}

\title{$DK/D\pi$ scattering and an exotic virtual bound state at the $SU(3)$ flavour symmetric point from lattice QCD}

\author[a]{J. Daniel E. Yeo,}
\author[a]{Christopher E. Thomas,}
\author[a]{David J. Wilson}
\author{\\(for the Hadron Spectrum Collaboration)}

\affiliation[a]{DAMTP, University of Cambridge, Centre for Mathematical Sciences, Wilberforce Road, Cambridge, CB3 0WA, UK}

\emailAdd{jdey2@cam.ac.uk}
\emailAdd{c.e.thomas@damtp.cam.ac.uk}
\emailAdd{d.j.wilson@damtp.cam.ac.uk}

\abstract{Elastic $S-$wave scattering of a charm meson with a light pseudoscalar meson in $J^P =0^+$ is investigated in the flavour $\bar{\mathbf{3}}$, $\mathbf{6}$ and $\overline{\mathbf{15}}$ sectors at the $SU(3)_f$ flavour point using lattice QCD, 
		 working on three volumes with $m_{\pi} \approx 700$ MeV.
		 Large bases of interpolating operators are employed to extract finite-volume spectra, 
		 which are subsequently used with the Lüscher method to provide constraints on infinite-volume scattering amplitudes.
		 Examining the singularities of the amplitudes, the $S-$wave amplitude in the flavour $\bar{\mathbf{3}}$ sector is found to contain a deeply bound state, strongly coupled to elastic threshold, 
		 corresponding to the $J^P = 0^+$ $D_{s0}^*(2317)$. 
                  In the exotic flavour $\mathbf{6}$ sector a virtual bound state is found at $\sqrt{s_{\rm{pole}}} = 2510 -  2610$ MeV, roughly $40-140$ MeV below threshold,
		 whereas the $\overline{\mathbf{15}}$ channel shows weak repulsion.
		 }

\begin{document} 
\maketitle

\section{Introduction}
\label{section:intro}
Over the last few decades the number of hadronic states observed experimentally has rapidly increased. 
Despite early success of the consistent quark model in describing these states, 
many of the more recent hadrons do not fit into the simple classification of conventional mesons and baryons,
and there are a number of such hadrons residing in the open and hidden charm and bottom sectors \cite{Chen:2016spr,Brambilla:2019esw,Chen:2022asf}.
In particular, the masses quoted by the Particle Data Group (PDG) review \cite{ParticleDataGroup:2022pth} for the charmed scalar mesons $D^*_0(2300)$ and $D^*_{s0}(2317)$ are very similar despite the latter containing a heavier strange quark.
While both the $D^*_0(2300)$ and $D^*_{s0}(2317)$ have quark model expectations of being broad resonances above $D\pi$ and $DK$ thresholds respectively \mbox{\cite{Godfrey:1985xj}},
this is only realised for the $D^*_0(2300)$~\mbox{\cite{Belle:2003nsh}} with the $D^*_{s0}(2317)$ instead seen as a narrow resonance below $DK$ threshold \mbox{\cite{BaBar:2003oey}}.

This situation has sparked much interest regarding the structure of the $D^*_{s0}(2317)$,
with ideas including a $DK$ molecule, 
a tetraquark or a conventional meson strongly coupled to the nearby $DK$ channel having been put forward \cite{Chen:2016spr}. 
To investigate this within QCD, as opposed to a QCD-inspired model, requires a non-perturbative approach. Lattice QCD, where spacetime is discretized on a lattice, provides such a first-principles method.
Some early lattice studies used only fermion-bilinear operators and neglected the presence of two-meson channels~\cite{Mohler:2011ke,Moir:2013ub,Perez-Rubio:2015zqb,Cichy:2016bci,Cheung:2016bym}, 
whereas others have also included a small number of meson-meson-like operators to extract near threshold scattering amplitudes~\cite{Mohler:2013rwa,Lang:2014yfa,Bali:2017pdv,Alexandrou:2019tmk}.
Chiral effective field theory methods and other approaches with lattice inputs have also been used ~\cite{MartinezTorres:2014kpc, Liu:2012zya,Guo:2018kno,Guo:2018tjx,Huang:2022cag,Lutz:2022enz,Gil-Dominguez:2023huq}.
More recently the Hadron Spectrum Collaboration (HadSpec) has investigated elastic $DK$ scattering using a large basis of $q\bar{q}$ and meson-meson operators, 
allowing for a robust determination of the energy dependence of the scattering amplitude~\cite{Cheung:2020mql}. 
There, the $D_{s0}^*$ is clearly seen as a bound state strongly coupled to the nearby scattering threshold. 

Despite appearing more in line with expectations, the $D^*_0(2300)$ has also sparked much discussion. 
There are fewer lattice studies dedicated to it; other than the previous HadSpec studies of~\cite{Moir:2016srx,Gayer:2021xzv},
only~\cite{Mohler:2012na} has investigated the  $D^*_0(2300)$ from $D\pi$ scattering.
Several chiral effective field theory studies have been performed with lattice inputs~\mbox{\cite{Guo:2018kno, Guo:2018tjx, Huang:2022cag, Lutz:2022enz}}.
With the $D^*_0(2300)$ seen as a broad resonance \cite{Belle:2003nsh},
care is needed as to how one defines its mass.
The PDG quotes the Breit-Wigner mass and, as has been pointed out in Refs.~\cite{vanBeveren:2003kd,vanBeveren:2006st,Gayer:2021xzv}, this does not necessarily coincide with the pole mass.
Furthermore, it has been argued that there is a two-pole structure in this sector with the interference between the two poles needed to explain the shape of the amplitude seen experimentally. 
UChPT suggests that the higher-energy pole can be identified with a pole in the flavour $\mathbf{6}$ in the $SU(3)_f$ flavour symmetry limit ~\cite{Kolomeitsev:2003ac, Albaladejo:2016lbb,Du:2020pui,Meissner:2020khl}.
Analysis using chiral effective field theory methods of more recent and precise LHCb data on 
$B_{(s)}$ decays \cite{Du:2017zvv,Du:2019oki} supports this hypothesis.
HadSpec studies of $D\pi$ scattering~\cite{Moir:2016srx,Gayer:2021xzv} only robustly determined a single pole in the elastic-scattering energy region near the $D^*_0(2300)$, though a second pole could be present at higher energies depending on the parameterisation used.
A separate analysis of the lattice data~\cite{Asokan:2022usm} suggests that a second pole is required if constraints from $SU(3)_f$ flavour symmetry are imposed on the $K-$matrix (with this symmetry only broken by
 different phase space factors),
and so clearly there are still questions to resolve.

In this study, we investigate scattering in the charm $J^P=0^+$ sector at the $SU(3)_f$ flavour point,
focusing on elastic $S$-wave scattering of pseudoscalar mesons.
The relevant scattering channels consist of a charm meson (flavour anti-triplet $\bar{\mathbf{3}}$) and
a light meson (flavour octet $\mathbf{8}$ or singlet $\mathbf{1}$), and
the resulting flavour decomposition is $\bar{\mathbf{3}} \oplus \mathbf{6} \oplus \overline{\mathbf{15}}$.
There are two main aims to this work. Firstly, in combination with previous HadSpec investigations, we will gain
a more complete picture of how phenomena in the charm sector depends on the light-quark mass. 
Secondly, working with $SU(3)_f$ symmetry where there is a separation into different flavour channels will allow us to determine whether a pole is present in the $\mathbf{6}$, compare with UChPT expectations and improve our understanding of the charm-sector physics.

The rest of this paper is organised as follows:
general methodology along with a discussion of how finite-volume energies are related to infinite-volume scattering amplitudes are presented in Section~\ref{section:methodology}, and 
then details of the lattice setup used are given in Section~\ref{section:lattice_details}.
In Section~\ref{section:scattering_analysis} the obtained finite-volume spectra, scattering amplitudes and poles are presented
for each of the flavour sectors in turn.
We interpret the results and compare with other work in Section \ref{section:disc}, and finish with a summary and outlook in Section \ref{section:concl_and_outlook}.

\section{Methodology} 
\label{section:methodology}
Stable hadrons and resonances can be rigorously defined in terms of poles of scattering amplitudes.
At low energy Quantum Chromodynamics (QCD) exhibits strongly-coupled dynamics and hence a non-perturbative method is needed to study hadronic interactions.
Lattice QCD provides one such method, replacing Minkowski spacetime with a finite-volume Euclidean lattice. 
Whilst scattering can not be investigated directly in this setup, 
the finite-volume formalism, pioneered by Lüscher and later extended,
allows one to indirectly relate finite-volume energy levels to scattering amplitudes of infinite-volume continuum QCD
-- see Ref.~\cite{Briceno:2017max} for a recent review.
The methodology we follow in this work is well established and we will briefly review the main details.

Calculations are performed on a cubic lattice in a finite cubic volume with periodic boundary conditions and,
due to the reduced symmetry compared to an infinite volume continuum, states are characterised by irreducible representations (irreps), $\Lambda$, of the remaining symmetry group
instead of by spin $J$. 
Momenta are restricted to take values $\vec{p} = \frac{2\pi}{L}\vec{n} \equiv [n_1n_2n_3]$, where $n_i \in \mathbb{Z}$ and $L$ is the spatial extent.
For a system with overall zero momentum, the relevant symmetry group is the (double cover of the) octahedral group and for non-zero overall momentum $\vec{P}$ it is the little group $\textrm{LG}(\vec{P})$.
We label irreps $[\vec{P}]\Lambda^{(P)}$, where parity $P$ is only a good quantum number for $\vec{P}=\vec{0}$.
Since we have the goal of obtaining information about QCD in an infinite-volume continuum, 
it is often useful to think about how irreps of the full continuous symmetry group decompose into direct sums of finite-volume irreps. 
Dealing with only mesons in this study, 
Table \ref{table:subductions} gives a summary of the decompositions for $\vec{P} = \vec{0}$ bosonic irreps from \cite{Johnson:1982yq}; those for $\vec{P} \neq \vec{0}$ can be found in~\cite{Moore:2005dw}.

The discrete spectrum on a finite-volume lattice can be obtained from two-point correlation functions using the variational method~\cite{Michael:1985ne,Luscher:1990ck, Blossier:2009kd}.
A matrix of correlation functions is computed, \smash{$C_{ij}(t) = \langle \Omega | \mathcal{O}_i(t)  \mathcal{O}^{\dagger}_j(0)| \Omega \rangle $}, 
where $\{\mathcal{O}_i\}$ is a basis of interpolating operators with the desired quantum numbers and appropriate structures, and a generalized eigenvalue problem (GEVP) is then solved,
\begin{equation}
	C_{ij}(t) \, v^{(\mathbf{n})}_j = \lambda^{(\mathbf{n})}(t,t_0) \, C_{ij}(t_0) \, v^{(\mathbf{n})}_j \, .
\end{equation}	
The eigenvalues can be used to obtain the finite-volume energies, $\{E_{\mathbf{n}}\}$, through $\lambda^{(\mathbf{n})}(t,t_0) \sim e^{-E_{\mathbf{n}}(t - t_0)}$, 
and the eigenvectors can be used to obtain operator overlaps, 
$Z^{\mathbf{n}}_i =\langle \mathbf{n} |\mathcal{O}^{\dagger}_i | \Omega \rangle$, and \emph{optimised operators}, 
${\Omega}^{\dagger}_{\mathbf{n}} \sim v^{(\mathbf{n})}_i {\mathcal{O}}^{\dagger}_i$ that give the optimal linear combination of the basis of operators to overlap onto energy eigenstate $| \mathbf{n} \rangle$.
Our implementation of the variational method is described in Ref.~\cite{Dudek:2010wm}.

In order to get a robust determination of the spectrum, an appropriate basis of operators must be chosen.
For states below the three-hadron threshold, it has been found that this can be achieved using a large basis of \emph{fermion-bilinear operators} and \emph{meson-meson operators}.
Appropriate fermion-bilinears for the charm sector have the form $\bar{\psi}_{\nu} \mathbf{\Gamma}\psi_{c}$, where $\mathbf{\Gamma}$ is a product of Dirac-matrices and gauge-covariant derivatives~\cite{Dudek:2010wm, Thomas:2011rh}, and $\nu \in \{u,d,s\}$ -- these operators transform under the $\bar{\mathbf{3}}$ irrep of $SU(3)_f$.
Meson-meson operators are formed from the product of
optimised operators for stable mesons, $\Omega_{\mathbb{M}_i}^{\dagger}(\vec{p}_i)$,
combined using lattice clebsch gordan coefficients, $CG$,
\begin{equation}
	\mathcal{O}^{\dagger}_{\mathbb{M}_1\mathbb{M}_2}(\vec{P}) \sim \sum_{\vec{p}_1, \vec{p}_2} 
	CG \ 
	\Omega_{\mathbb{M}_1}^{\dagger}(\vec{p}_1)\
	\Omega_{\mathbb{M}_2}^{\dagger}(\vec{p}_2) \, ,
\end{equation}
with the sum over all $\vec{p}_i$ related by finite-volume rotations 
under the constraint $\vec{p}_1 + \vec{p}_2 = \vec{P}$ -- see Ref.~\cite{Dudek:2012gj} for details.
The meson-meson operators used in this work have $\mathbb{M}_1$ a $D^{(\ast)}$ meson (in the $SU(3)_f$ flavour $\mathbf{\bar{3}}$ irrep) and $\mathbb{M}_2$
a light meson (in the flavour $\mathbf{8}$ or $\mathbf{1}$ irrep), 
yielding the flavour decomposition $\bar{\mathbf{3}} \oplus \mathbf{6} \oplus \overline{\mathbf{15}}$.
We do not include compact tetraquark-like operators $\mathcal{O}\sim[q^Tq][\bar{q}^T\bar{q}]$ in the basis -- studies in the charm sector have shown that the obtained spectrum has negligible sensitivity to the inclusion of such operators~\cite{Cheung:2017tnt, Alexandrou:2019tmk}.
The fermion-bilinear and meson-meson operators in our operator basis are appropriately projected into the finite-volume irrep of interest.

\begin{table}[t!]
	\centering
	  \begin{tabular}{|c|c|}
		\hline 
		\rule{0pt}{12pt} $J^P$&    $\Lambda^P(\text{dim}[\Lambda])$  \\
		\hline 
		\rule{0pt}{11pt} $0^{\pm}$ &  $A_1^{\pm}(1)$  														   \\[3pt]
		$1^{\pm}$				   &  $T_1^{\pm}(3)$ 														   \\[3pt]
		$2^{\pm}$    			   &  $T_2^{\pm}(3) \oplus E^{\pm}(2)$ 										   \\[3pt]
		$3^{\pm}$ 				   &  $T_1^{\pm}(3) \oplus T_2^{\pm}(3) \oplus A_2^{\pm}(1)$    			   \\[3pt]
		$4^{\pm}$ 				   &  $A_1^{\pm}(1) \oplus T_1^{\pm}(3) \oplus T_2^{\pm}(3) \oplus E^{\pm}(2)$ \\[3pt]
		\vdots 					   & \vdots						     										   \\[3pt]

	  \hline
	\end{tabular}
	  \caption{\label{table:subductions}Infinite-volume spin decompositions into at-rest finite-volume irreps, $[000]\Lambda^P$ \cite{Johnson:1982yq}.}
\end{table}
\vspace{0.5cm}

Once the finite-volume spectrum has been obtained, 
it can be used to determine infinite-volume scattering amplitudes 
using the approach developed by Lüscher~\cite{Luscher:1990ux} and subsequent extensions~\cite{Luscher:1986pf, Fu:2011xz, Luscher:1991cf, Kim:2005gf, Christ:2005gi, Leskovec:2012gb, Hansen:2012tf, Briceno:2014oea}.
The relation between scattering amplitudes and finite-volume spectra is given by the quantisation condition,
\begin{equation}
	\label{eq:luscher}
	\det \Big{[} \mathds{1} + i \rho (s) t(s)\big{(} \mathds{1} + i \overline{\mathbf{\mathcal{M}}}^{\Lambda(\vec{P})}(s, L)\big{)}\Big{]}  = 0 \, ,
\end{equation}
where the determinant is over the space of hadron-hadron channels and partial waves, $t(s)$ is the (partial-wave projected) $t$-matrix, 
$\rho(s) = \text{diag}(\frac{2k_1}{\sqrt[]{s}}, \frac{2k_2}{\sqrt[]{s}}, ...)$ is a diagonal matrix of phase-space factors with $k_i$ the scattering momentum in the centre-of-mass (CM) frame for channel $i$,
and \smash{$\overline{\mathcal{M}}$} is a volume, energy and finite-volume irrep dependent matrix of known functions~\cite{Briceno:2014oea}.
The expression, valid up to exponentially suppressed finite-volume terms which are assumed to be negligible, is a function of a single Mandelstam variable $s=E^2_{\sf cm}$, where $E_{\sf cm}$ is the energy in the CM frame, 
and spatial length $L$.
The solutions of this quantisation condition are the finite-volume energy levels $\{E_{{\sf cm}, \mathbf{n}}\}$ in irrep $[\vec{P}]\Lambda$.
With the reduced symmetry from working in a finite cubic volume, scattering amplitudes of many different $J$ can contribute to the same $[\vec{P}]\Lambda$ irrep's spectrum, 
and therefore, in principle, must be included together when finding solutions to the quantisation condition. 
However, we can often focus on just a small number of scattering channels when only a few channels are kinematically open or when some channels are suppressed.
For example, unless enhanced by a nearby resonance or bound state, channels involving higher orbital angular momentum are suppressed by powers of the scattering momentum close to threshold.

For the case of elastic $S-$wave scattering of pseudoscalar mesons\footnote{Where elastic in this context means only one channel is kinematically open.} -- the focus of this study -- 
 expressing the $t-$matrix in terms of a phase shift $\delta(s)$ via $t(s)= \frac{1}{\rho(s)} e^{i \delta(s)} \sin {\delta}(s)$,
the quantisation condition simplifies to the compact expression,
\begin{equation}
	\label{eq:elastic_luscher}
	k \cot \delta(s) = - k \cot \phi (s,L)
\end{equation} 
where the explicit form of $\phi(s,L)$ can be found in Ref.~\cite{Briceno:2017max}.
Since the obtained spectra yield a finite set of constraints via Eq.~\ref{eq:luscher}, 
we use a $\chi^2$ approach to fit energy-dependent parameterisations of the $t-$matrix to the data~\cite{Dudek:2012xn, Woss:2020cmp}, 
with many different parameterisations considered to avoid potential bias.
One useful form is the effective range parameterisation which for $S$-wave scattering is
\begin{equation}
	k\cot\delta(k) = \frac{1}{a} + \frac{1}{2}rk^2 +P_2 k^4 + \mathcal{O}(k^6)\, . 
\end{equation}
$K$-matrix parameterisations are
a more general class of parameterisations that can also deal with coupled hadron-hadron channels and/or partial waves\footnote{Despite a general focus on elastic $S-$wave scattering, 
coupled-channel scattering is needed in the analysis of the flavour
$\mathbf{6}$ sector (for more details see Section~\ref{section:6f}).}
-- these are defined by
\begin{equation}
	(t^{-1})_{ij} = \frac{1}{(2k_i)^{\ell_i}}(K^{-1})_{ij}\frac{1}{(2k_j)^{\ell_j}} + I_{ij}\,\, ,
\end{equation}
with $s$-channel unitarity satisfied if $\text{Im}[I(s)_{ij}]= - \rho_i(s)\delta_{ij}$ above threshold and zero below, and $K$ a real-valued matrix for real $s$.
We will consider parameterisations where $K$ is a polynomial, a ratio of polynomials, and a polynomial with an explicit pole term  $K\sim1/(s-m^2_0$).
There is freedom in the choice of $I$, and we utilise two in this study: the \emph{simple phase-space prescription} \smash{$I(s)_{ij}= -i \rho_i(s)\delta_{ij}$}
and the \emph{Chew-Mandelstam prescription} \cite{PhysRev.119.467} where the real part of $I(s)_{ij}$ is related to its known imaginary part via a dispersive integral -- our implementation is described in Ref.~\cite{Wilson:2014cna}.
When using the Chew-Mandelstam prescription a subtraction point must be specified.
For $K$-matrices with an explicit pole term we set $I(m^2_{0})=0$, and otherwise $I(s_{\text{thr}})=0$.

\vspace{0.5cm}

Once reasonable parameterisations have been determined, the scattering amplitudes can be analytically continued to the complex $s$-plane.
With unitarity enforcing a square-root singularity at each kinematic threshold, 
for elastic scattering the $s$-plane splits into two Riemann sheets, denoted by the sign of $\text{Im}[k(s)]$. 
\emph{Bound states} are found on the real axis of the `physical sheet' ($\text{Im}[k(s)]>0$) below threshold, and 
\emph{virtual bound states} and \emph{resonances} are found on the `unphysical sheet' ($\text{Im}[k(s)]<0$).
By determining the analytic structure of the parameterisations, poles can be found and hadrons identified.
Near a pole, the $t-$matrix takes the factorised form 
\begin{equation}
	t_{ij} \sim \frac{c_ic_j}{s_{\text{pole}}-s}
\end{equation}
where  $c_i$ gives  the pole's coupling to scattering channel $i$.

\begin{table}
	\centering
	\resizebox{0.58\columnwidth}{!}{%
		\centering
		\begin{tabular}{|cccccc|}
			\hline
			\rule{0pt}{11pt} $(L/a_s)^3\times (T/a_t)$    & $N_{\text{cfgs}}$   & $N_{\text{tsrcs}}$  & $N_{\text{vecs}}$    & \footnotesize \makecell{Dispersion \\ Analysis}   & \footnotesize \makecell{Scattering \\ Analysis} \\[3pt]
			\hline
			\hline
			\rule{0pt}{13pt} $12^3 \times 96$      	     & 219	& 4 & 48   	& \checkmark     &    		     \\[2pt]
			$14^3 \times 128$      	  				     & 397	& 4 & 64       & \checkmark     &  			 \\[2pt]
			$16^3 \times 128$      	     			     & 533	& 4 & 64       & \checkmark     &  \checkmark   \\[2pt]
			$18^3 \times 128$      	     				 & 358	& 4 & 96       & \checkmark     &   		     \\[2pt]
			$20^3 \times 128$      	    				 & 503	& 4 & 128      & \checkmark     &  \checkmark   \\[2pt]
			$24^3 \times 128$      	   				     & 607	& 4 & 160      & \checkmark     &  \checkmark   \\[2pt]
			\hline
		\end{tabular}
		}
		\caption{Lattices used, with spatial and temporal extents, $L$ and $T$,
				 number of configurations $N_{\text{cfgs}}$, number of time sources $N_{\text{tsrcs}}$ and number of distillation vectors $N_{\text{vecs}}$.
				 Ticks show which stage of the analysis each volume was used in.}
		\label{table:ensembles}
\end{table}

Whilst the Lüscher formalism does capture the correct analytic structure of $t(s)$ from $s$-channel
scattering, it doesn't capture the kinematic or dynamical `left-hand cuts' (LHC) generated from $t$-channel and $u$-channel crossing symmetry or particle exchange that appear in partial-wave projected $t$-matrices. 
This potentially causes a problem when applying the quantisation condition to finite-volume energies far below scattering threshold.
Recent work has been done to incorporate some of these LHC effects into the formalism by explicity adding one-particle exchange cross-channel contributions
to the usual two-particle finite-volume formalism \cite{Raposo:2023oru} or using three-particle formalism of \cite{Hansen:2014eka, Hansen:2015zga} and extracting the relevant 2-to-2 amplitudes as described in \cite{Hansen:2024ffk},
but these have not yet been demonstrated in practice.
The closest LHC is lower in energy than the finite-volume energies used in this study, and we leave the investigation of its effects for future projects.\footnote{
	The nearest LHC, generated by a single $\sigma$ particle t-channel exchange, appears at an energy \mbox{$a_t E = 0.5180(9)$}.}

\section{Lattice setup} 
\label{section:lattice_details}

\subsection{Lattice details}

\begin{figure}[tbp]
	\includegraphics[width=\linewidth]{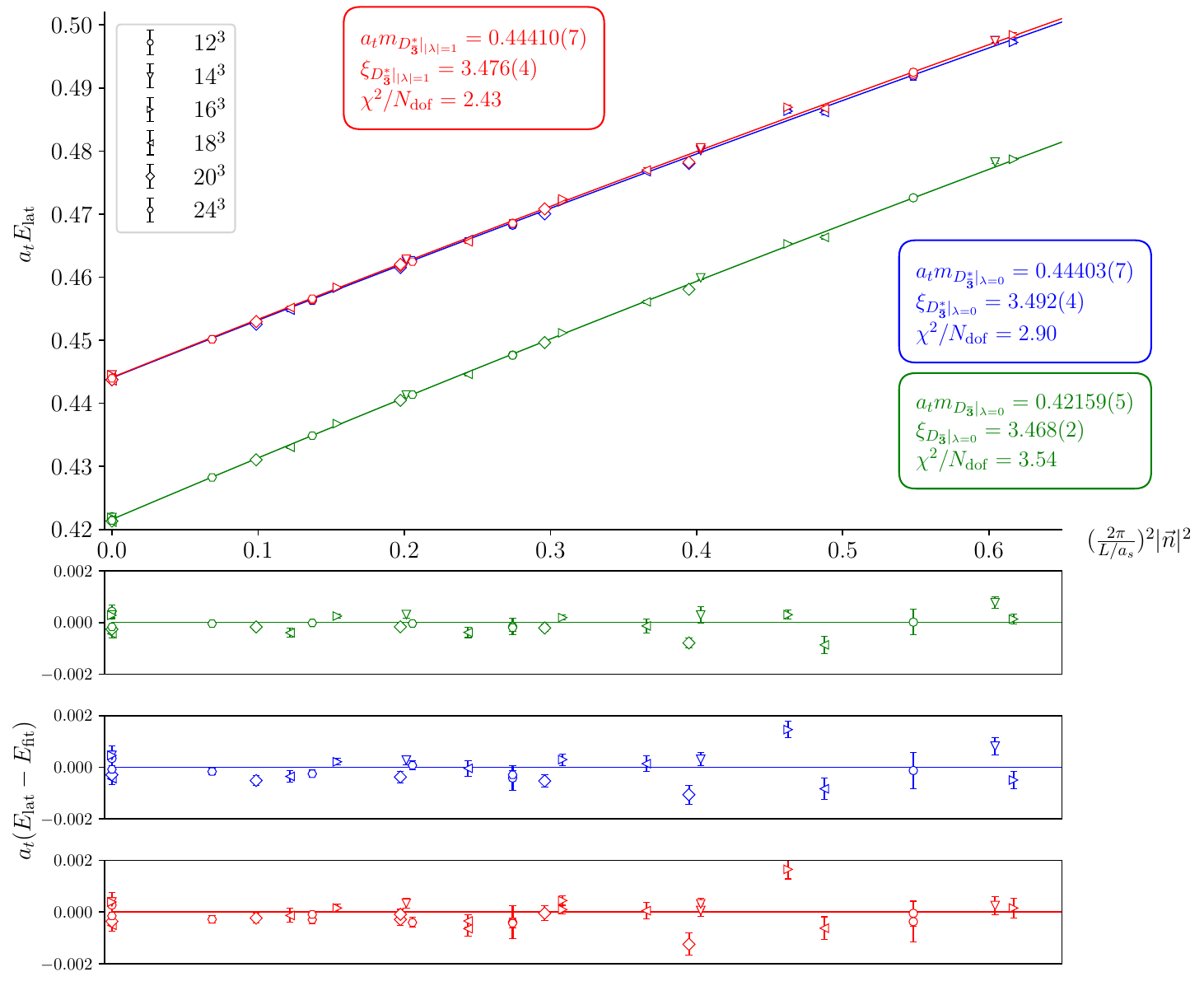}
	\caption{Top: Finite-volume energies (points) corresponding to the $\D$ (green),  $\Dst|_{\lambda =0}$ (blue) and $\Dst|_{|\lambda| =\pm1}$ (red) as a function of momentum squared 
			 $(pa_s)^2 = (\frac{2\pi}{L / a_s})^2|\vec{n}|^2$. 
			 Dispersion fits to these energies (lines) are presented
			 with their associated masses, anisotropies and \chisq. 
			 Lower panels: Deviation of obtained lattice energies $(E_\text{lat})$ when compared to the dispersion fits $(E_\text{fit})$.
			}\label{fig:Dmesons_disp}
\end{figure}

For this investigation, anisotropic lattices were used
where the temporal lattice spacing $a_t$ is finer than the spatial lattice spacing $a_s$, 
characterised by the anisotropy $\xi = \frac{a_s}{a_t}\approx 3.5$.
The gauge sector is described by a tree-level Symanzik-improved anisotropic action and 
the fermion sector by a tadpole-improved anisotropic Sheikholeslami-Wohlert (Clover) action
with spatial stout-smearing~\cite{Edwards:2008ja}.
There are three degenerate dynamical `light' quarks and one quenched charm quark with bare parameters chosen such that,
after appropriate scale setting, they yield the physical strange and charm quark masses respectively~\cite{HadronSpectrum:2008xlg, HadronSpectrum:2012gic}.
Using $N_{\text{cfg}}$ gauge configurations, the uncertainties are propagated via jackknife resampling. 
We utilize the distillation framework, allowing for 
fermionic field smearing and efficient computation of the correlation functions involving large bases of fermion-bilinear and meson-meson operators~\cite{HadronSpectrum:2009krc}.
With these parameters we have 6 different lattice volumes as summarised in Table~\ref{table:ensembles},
but note that only the $L/a_s =16,20,24$ volumes were used for the main analysis in Section~\ref{section:scattering_analysis}.
The spatial and temporal extents are large enough for the finite-volume corrections for stable hadrons and finite temporal effects expected to be negligible, with $m_{\pi}L \approx 6$ and $m_{\pi}T \approx 14$ on the smallest volume lattice.
To present the results in physical units, 
we determine $a_t$ by requiring the $\Omega$-baryon mass measured on the $L/a_s=16$ volume to be equal to
the experimental mass quoted by the PDG, 
leading to $a_t^{-1}  = 4655$ MeV~\cite{Woss:2018irj}. 
Next we discuss the relevant stable mesons that can contribute to the scattering amplitudes of interest.

\subsection{Stable mesons at the flavour symmetric point} 
\label{section:stable_mesons}

In this study we are interested in the scattering of a charmed meson and a light meson close to the lowest kinematic threshold. 
The light mesons become exact flavour octets and singlets in the limit of $SU(3)_f$ flavour symmetry.  
The pseudoscalar flavour octet, singlet are labelled as $\etaoct$, $\etas$
and the vector flavour octet, singlet as $\omoct$, $\oms$, 
with the $\pi$ contained in $\etaoct$ and the $\rho$ contained in $\omoct$.
The scalar flavour singlet is labelled as $\fzs$ and contains the $\sigma$ meson. 
Furthermore, in the $SU(3)_f$ flavour limit, 
the pseudoscalar charmed mesons $D(c\bar{l})$, $D_s(c\bar{s})$ are combined into a single pseudoscalar flavour $\bar{\mathbf{3}}$ multiplet, $\D$, 
and the vector charmed mesons $D^*(c\bar{l})$, $D_s^*(c\bar{s})$ are packaged into a single vector flavour $\bar{\mathbf{3}}$ multiplet, $\Dst$.
\begin{table}[t!]
	\centering
	\parbox{1\linewidth}{
		\centering
		\begin{tabular}{|ccccc|}
			\hline
			\rule{0pt}{12pt}Meson         & $a_tm$        & $m/$MeV     & Flavour                & $J^P$ \\
			\hline
			\hline 
			$\etaoct$      			     & 0.1478(1)	 & 688.0(5)    & $\boldsymbol{8}$       & $0^-$  \\
			$\etas$         			 & 0.2017(11) 	 & 938(5)      & $\boldsymbol{1}$       & $0^-$  \\
			$\omoct$     	             & 0.2154(2)     & 1002.7(9)   & $\boldsymbol{8}$       & $1^-$  \\
			$\oms$        				 & 0.2174(3)     & 1012(1)     & $\boldsymbol{1}$       & $1^-$  \\
			$\fzs$ 						 & 0.2007(18)    & 934(8)      & $\boldsymbol{1}$       & $0^+$  \\
			$\D$       					 & 0.42159(5)   & 1962.5(2)   & $\boldsymbol{\bar{3}}$ & $0^-$  \\
			$\Dst$			             & 0.44407(11)   & 2067.1(5)     & $\boldsymbol{\bar{3}}$ & $1^-$  \\
			\hline

		\end{tabular}
		\caption{Relevant stable hadrons and their masses, 
				 with the light mesons taken from Refs.~\cite{Woss:2018irj, Woss:2020ayi}.
				}
	\label{table:meson_mass}
	}
\end{table}

To study the momentum dependence of the relevant stable charmed mesons,
large bases of fermion bilinear operators were employed to extract the ground state in the relevant finite-volume irreps with total momentum $\vec{P}= \frac{2\pi}{L}\vec{n}$ for $|\vec{n}|^2 \leq 2$ on the $12^3$, $|\vec{n}|^2 \leq 3$ on the $14^3$ and $|\vec{n}|^2 \leq 4$ on the $\{16^3, 18^3, 20^3, 24^3\}$ lattices.
Fig.~\ref{fig:Dmesons_disp} shows the extracted energies as a function of momentum. 
Fitting to the relativistic dispersion relation for stable hadrons on an anisotropic lattice,
\begin{equation}
	\label{eq:dispersion}
	(a_tE_{\vec{n}})^2 = (a_tm)^2 +\frac{1}{\xi^2}\Big{(} \frac{2 \pi}{L/a_s}|\vec{n}|\Big{)}^2 \, ,
\end{equation}
the masses and anisotropies were determined 
and are presented on the figure along with the associated \chisq.
For the \Dst\ we perform separate fits for the two different helicity magnitudes, $|\lambda| = 0$ and $1$.
In the lower panels of the figure deviations of the obtained lattice energies from the dispersion fits are shown.
Despite energy levels appearing to follow the expected trend well, 
the resulting \chisq\ are relatively poor.
In absolute terms these deviations are small\footnote{Taking the extreme cases of $|a_t(E_{\text{lat}}-E_{\text{fit}})|= 0.002$ with a rest mass 0.444, this is a $\sim 0.5\%$ deviation.}
but they are statistically significant due to the high precision of the energy levels, 
revealing systematic errors.
Similar effects have been observed in previous HadSpec studies \cite{Cheung:2020mql, Wilson:2023anv}, 
and more general discussions of the systematic uncertainties in the charm sector using our lattice setup, for example a smaller $J/\psi-\eta_c$ mass splitting when compared to experiment, can be found in  \cite{HadronSpectrum:2012gic, Cheung:2016bym, Moir:2013ub}.

As in Ref.~\cite{Cheung:2020mql}, to account for some of these systematics
we add additional systematic uncertainty to the finite-volume energy levels,
\begin{equation}
 \delta E = \sqrt[]{ \delta E^2_{\rm{stat}} + \delta E^2_{\rm{add}}} \, ,
\end{equation}
with a value of $a_t\delta E_{\rm{add}}=0.0005$ chosen based roughly on the spread of obtained masses of the $D$ mesons across the different volumes
(see the at-rest energies on the lower panels of Fig~\ref{fig:Dmesons_disp}).
This is typically a small increase in error.\footnote{With the most extreme cases resulting in a doubling of error, see Fig.~\ref{fig:6f_spectra} for examples.}

\begin{table}
	\centering
	\begin{minipage}{.8\linewidth}
	\centering
	  \begin{tabular}{|cc|cc|cc|}
		\hline
		\rule{0pt}{11pt} Channel &    $a_tE_{\sf thr}$  & Channel &    $a_tE_{\sf thr}$ & Channel &    $a_tE_{\sf thr}$ \\[3pt] 
		\hline
		\hline
		\rule{0pt}{11pt} \D\etaoct   &  0.56939(11)     & \D\oms			  	   &  0.63899(30)	& $\D\etaoct\etaoct$ & 	 0.71719(21) \\[3pt]
		\rule{0pt}{0pt}\Dst\etaoct	 &  0.59187(15) 	& \Dst\fzs				   &  0.64477(180)	&& 	   \\[3pt]
		\D\fzs				  	     &  0.62229(180)	& \Dst\etas    		  	   &  0.64577(111)	&&	   \\[3pt]				
		\D\etas    		  	         &  0.62329(110)    & \Dst\omoct			   &  0.65947(23)	&&	   \\[3pt]	
		\D\omoct			  	     &  0.63699(21)     & \Dst\oms			  	   &  0.66147(32)	&&	   \\[3pt]	
		\hline
	\end{tabular}
\end{minipage}
	  \caption{Relevant kinematic thresholds ordered by energy.}
	  \label{table:thresholds}
\end{table} 

Masses of relevant stable mesons are summarised in Table \ref{table:meson_mass},
with those for light mesons taken from \cite{Woss:2018irj, Woss:2020ayi}.
For vector mesons the mass used is a mean of the masses from fits to the $|\lambda| = 0$ and $1$ components
with an uncertainty encompassing the uncertainties on the masses from the individual fits.
Using these masses we determine the relevant scattering channels and their associated kinematic thresholds which are summarised in Table~\ref{table:thresholds}.
For the anisotropy to use in the scattering analysis, we take the mean of the anisotropies obtained for the 
$\etaoct$ ($3.446 \pm 0.003$ from Ref.~\cite{Woss:2018irj}), the $\D$ meson, and the $|\lambda| = 0$ and $1$ components of the $\Dst$. 
This gives a value of $\xi = 3.471\pm{0.028}$, where the uncertainty has been chosen to encompass the uncertainties of the individual hadron anisotropies.

\section{Results} 
\label{section:scattering_analysis}

We now discuss results for the three flavour sectors in turn.
For each sector, the infinite-volume scattering channel contributions,
the obtained spectra for each irrep, the subsequent scattering analysis 
and any poles in the amplitudes are discussed.

\subsection{Flavour \texorpdfstring{$\bar{\mathbf{3}}$}{Lg} Sector}

\subsubsection{Finite-volume spectrum}
Our focus is on $J^P = 0^+$ elastic \D\etaoct\ scattering close to threshold.
From Table~\ref{table:subductions} we note that the $[000]A_1^+$ irrep contains information,
via the Lüscher quantisation condition, on the $J^P=0^+$ sector $t$-matrix 
along with $t$-matrices from higher spin sectors ($J^P = 4^+,...$) that we are not directly interested in.
The associated scattering channels are summarised in Table~\ref{table:3bar_subductions}, 
with contributions from both $D^{(*)}$-singlet and $D^{(*)}$-octet
since both can carry flavour quantum numbers of the $\bar{\mathbf{3}}$ irrep.
Channels are ordered by their kinematic thresholds
and labelled by partial wave \SLJ{2S+1}{\ell}{J} spectroscopic notation,
where $S$ is the total intrinsic spin, $\ell$ is the orbital angular momentum and $J$ is the total angular momentum,
with contributions from $G$-wave ($\ell=4$) or higher and $J>4$ omitted from the table.
Noting that the first inelastic channel, $S$-wave \D\etas, opens at $\sim0.623$,
we stay well below this energy when performing the analysis of elastic $S$-wave \D\etaoct\ scattering.

\begin{table}
	\centering
	\resizebox{0.5\columnwidth}{!}{%
		\begin{tabular}{|cccl|}
			\hline
			\rule{0pt}{12pt} $[000]\Lambda^P$ 	& $J^P$  & Channels  	& \SLJ{2S+1}{\ell}{J} \\
			\hline
			\hline
			\rule{0pt}{12pt} $[000]A_1^+$ 	& $ 0^+,\ \grey{4^+,...}$	& \D\etaoct  		& \SLJ{1}{S}{0} \\[3pt]
										    & 							& \D\etas  		& \grey{ \SLJ{1}{S}{0} }\\[3pt]
										    & 							& \Dst\fzs  		& \grey{\SLJ{3}{P}{0}, \SLJ{3}{F}{4}} \\[3pt]
										    & 							& \vdots 	        & \vdots  \\[3pt]
			\hline
	
		\end{tabular}
			}
		\caption{Infinite-volume scattering channel contributions into finite-volume irrep $[000]A_1^+$ for the flavour $\bar{\mathbf{3}}$ sector, 
				 with $G$-wave or higher and $J>4$ contributions omitted from the table.
				 Channels not included in the scattering analysis are in grey.
				 Channels are ordered by increasing kinematic threshold.
				 }
		\label{table:3bar_subductions}
\end{table}

\begin{figure}[tbp]
	\centering
	\minipage{0.8\textwidth}
	  \includegraphics[width=\linewidth]{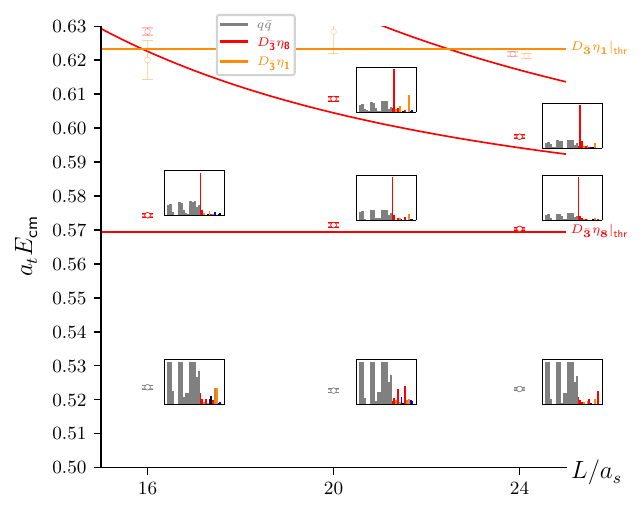}
	\endminipage
	  \caption{Finite-volume spectrum for the $[000]A^+_1$ irrep in the flavour $\bar{\mathbf{3}}$ sector (points) alongside non-interacting meson-meson energies (curves).
			   Each point is shown with both statistical uncertainty and additional systematic uncertainty (see Section \ref{section:lattice_details}).
			   Some points are horizontally displaced for clarity.
			   Curves are coloured by hadron pair and levels are coloured by dominant operator overlaps, with the key given in the figure.  
			   Energy levels not included in the scattering analysis are faded out.
			   For levels used in the scattering analysis, normalised operator overlaps are presented as histogram plots
			   as described in the text with meson-meson operators ordered by non-interacting energy. 
			   }\label{fig:3bar_spectra}
\end{figure}

To obtain the spectrum in the $[000]A_1^+$ irrep, 
a large basis of fermion-bilinear and meson-meson interpolating operators were employed to compute the matrix of correlation functions
-- a list of the operators is given in Table~\ref{table:3bar_interpolating_list} in Appendix~\ref{appendix:interpolating_list}.
The variational method was then used to extract the spectrum as discussed in Section~\ref{section:methodology}
and the obtained spectrum for each volume is presented in Fig.~\ref{fig:3bar_spectra}.
Also plotted are non-interacting meson-meson energy curves,
i.e. the expected meson-meson energies in the absence of meson-meson interactions,
coloured according to hadron pair.
Furthermore, we present normalised operator overlaps 
$\tilde{Z}^{\mathbf{n}}_i =\langle \mathbf{n} |\mathcal{O}^{\dagger}_i | \Omega \rangle / \text{max}_{\mathbf{m}}\{|Z^{\mathbf{m}}_i|\}$,
with a normalisation such that for a given operator its maximum overlap with any state is one.
We observe that the lowest finite-volume energy level on each volume is far below threshold and overlaps predominantly with a selection of $q\bar{q}$ operators.
Each of the other levels
 has a dominant overlap with a single meson-meson operator and lies close to but slightly above the corresponding non-interacting meson-meson energy,
and the points are coloured accordingly.
These features are usually an indication of a bound state in the corresponding infinite-volume scattering matrix \cite{Luscher:1991cf}.
Staying well below the first inelastic threshold, the 8 levels below $a_t E = 0.61$ are used
for elastic $S$-wave \D\etaoct\ scattering analysis. 
Levels not included in the analysis are faded out in the spectrum plot.
It was found that using this single at-rest irrep for 3 different volumes $(L/a_s = 16, 20, 24)$ gave enough constraints on the scattering amplitude.
Numerical values of the finite-volume energies with their uncertainties and correlations are provided in the supplementary material.

\subsubsection{Scattering amplitudes and poles}
\label{section:scalar_analysis}

\begin{figure}
	\centering
	\minipage{0.95\textwidth}
	\includegraphics[width=\linewidth]{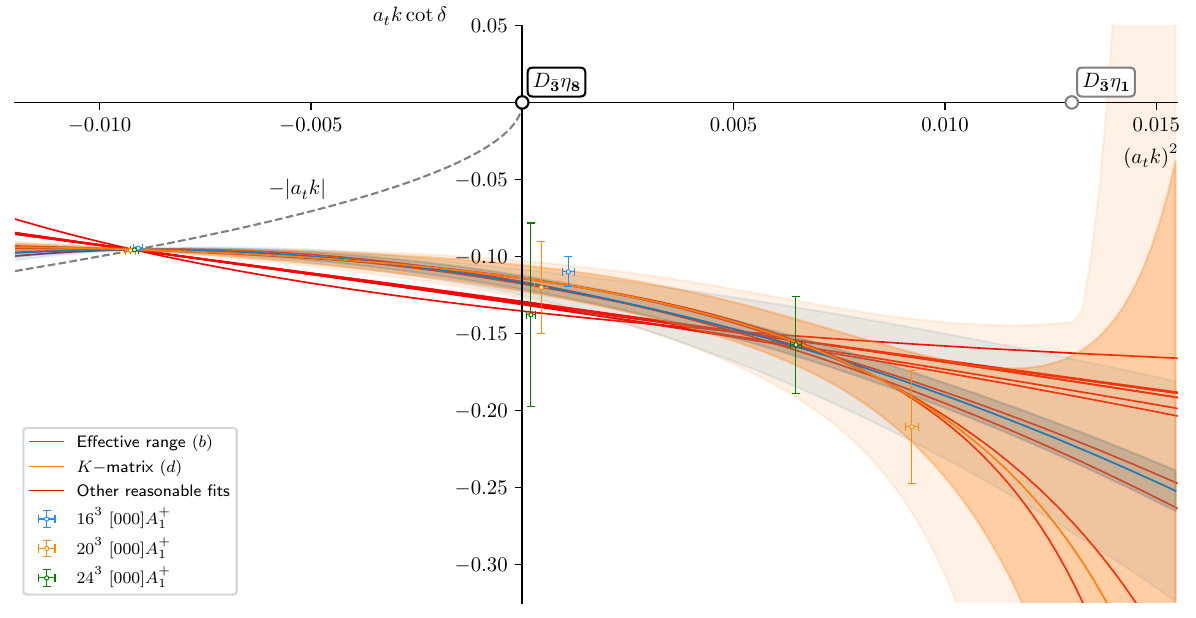}
	\endminipage
	\caption{Elastic $S$-wave $\D\etaoct$ scattering phase-shift $\delta(k)$, in the flavour $\bar{\mathbf{3}}$ sector, plotted as $k\cot \delta(k)$. 
			 Points are the result of using Eq.~\ref{eq:elastic_luscher} for each energy and are coloured according to volume.
			 The two reference parameterisations, effective range $(b)$ and $K$-matrix $(d)$, are each shown with a central value, statistical uncertainty (inner band)  
			 and uncertainty from mass and anisotropy variations as described in the text (outer band).
			 Central values of other reasonable fits are shown as red curves.			 
			 Circles on the horizontal axis indicate kinematic thresholds.
			}\label{fig:3bar_kcot}
\end{figure}

\begin{table}
	\centering
	\resizebox{0.85\columnwidth}{!}{%
		\begin{tabular}{ lcc}
			\hline
			Parameterisation  				  & \multicolumn{1}{c}{$\chi^2 / N_{\rm{dof}} \ $ } & \multicolumn{1}{c}{$a_t\sqrt{s_{\rm{pole}}}$}\\
			\hline
			  
			Effective range                                                    					     &                                    			    &                                            \\[5pt] 
			$(a)\ k \cot \delta(k) = \frac{1}{a} + \frac{1}{2}rk^2 $         					     & $\frac{7.04}{8-2}= 1.17 $                       & 0.52319(72)                                 \\[8pt]
			$\bm{(b)\ k} \textbf{ cot} \bm{\delta(k) = \frac{1}{a} + \frac{1}{2}rk^2  + P_2k^4}$ 	 & $\bm{\frac{3.79}{8-3}= 0.76} $                  & $\bm{0.52338(34)}$                             \\[8pt]
			\hline
			$K$-matrix with Chew-Mandelstam phase-space                         &                                        		  &                                            \\[5pt] 
			$(c)\ K(s) = \frac{g_0^2}{s-m^2_0}$                                 & $\frac{6.73}{8-2} = 1.12$                       & 0.52320(33)                                \\[8pt]
			$\bm{(d)\ K(s) = \frac{g_0^2}{s-m^2_0} +\gamma_0$ }                 & $\bm{\frac{3.04}{8-3} = 0.61}$                  & $\bm{0.52337(32)}$                          \\[8pt]
			$(e)\ K(s) = \frac{g_0^2}{s-m^2_0} +\gamma_1s$                & $\frac{2.94}{8-3} = 0.59$                       & 0.52336(36)                                \\[8pt]
			$(f)\ K(s) = \frac{(g_1s)^2}{s-m^2_0}+\gamma_0$               & $\frac{3.12}{8-3} = 0.62$                       & 0.52337(37)                                \\[8pt]
			$(g)\ K^{-1}(s) = c_0 + c_1 s$                        		& $\frac{6.73}{8-2} = 1.12$                       & 0.52320(33)                                \\[8pt]
			$(h)\ K^{-1}(s) = c_0 + c_1 s +c_2 s^2$                 & $\frac{3.92}{8-3} = 0.78 $		              & 0.52341(36)                                \\[8pt]
			\hline 
 
 			$K$-matrix with simple phase-space                 	        &                                                 &                                            \\[5pt]
			$(i)\ K(s) = \frac{g_0^2}{s-m^2_0}$                                 & $\frac{7.06}{8-2} = 1.18$      				    & 0.52319(35)                                \\[8pt]
			$(j)\ K(s) = \frac{g_0^2}{s-m^2_0} +\gamma_0$                       & $\frac{6.69}{8-3} = 1.34$       			        & 0.52317(36)                                \\[8pt]
			$(k)\ K(s) = \frac{g_0^2}{s-m^2_0} +\gamma_1s$                & $\frac{6.52}{8-3} = 1.30$         			    & 0.52317(35)                                \\[8pt]
			$(l)\ K(s) = \frac{(g_1s)^2}{s-m^2_0}+\gamma_0$               & $\frac{9.88}{8-3} = 1.98$                         & 0.52314(32) 					             \\[8pt] 
			$(m)\ K^{-1}(s) = c_0 + c_1 s$                        	    & $\frac{7.06}{8-2} = 1.18$                         & 0.52319(33)                                \\[8pt]
			$(n)\ K^{-1}(s) = c_0 + c_1 s +c_2 s^2$                 & $\frac{3.85}{8-3} = 0.77$		                    & 0.52337(32)                                \\[8pt]
			\hline  
		\end{tabular}
	}
	\caption{List of parameterisations, \chisq\ and associated physical-sheet pole locations for elastic
			 $S$-wave \D\etaoct\ scattering in the flavour $\bar{\mathbf{3}}$ sector. 
			 Reference parameterisations are in bold.
			 }\label{table:3bar_fits} 
\end{table}

The Lüscher quantisation condition was used to relate the infinite-volume $t$-matrix to finite-volume energies. 
As discussed in Section~\ref{section:methodology}, 
for elastic scattering there is a one-to-one relation between energy levels and the scattering matrix sampled at those energies.
The result of applying Eq.~\ref{eq:elastic_luscher} to the each of the finite-volume energies is given by the points in Fig.~\ref{fig:3bar_kcot} where the scattering phase shift is plotted as $k \cot \delta(k)$.

A number of effective range and $K$-matix parameterisations for the scattering matrix were considered and fit to the data as described in Section~\ref{section:methodology}.
The parameterisations are summarised in Table~\ref{table:3bar_fits} with full results of the fits given in the supplementary material.
To illustrate the results we choose two reference parameterisations: the second order effective range expansion $(b)$
and a $K$-matrix parameterisation $(d)$,
highlighted in bold in the table.
To assess the systematic uncertainty from the uncertainty on the scattering hadron masses and anisotropy, 
we perform \emph{mass-anisotropy variations} on the reference parameterisations where $1\sigma$ variations on the hadron masses and
on the anisotropy are considered.
For the effective range $(b)$ we obtain the fit parameters,
\begin{align} 
	\label{eq:3bar_effrange_fit_params}
	\begin{split}
		a   = &\ (-8.54 \pm 0.31 \pm 0.64) \cdot a_t \\ 
		r   = &\ (-9.46 \pm 0.86 \pm 4.07 \cdot a_t\\   
		P_2 = &\ (-260 \pm 40 \pm 158) \cdot a^{3}_t\\ 	
	\end{split} \
        \quad
	\begin{bmatrix} 
		\ 1\ \ & 0.46 & 0.34\\ 
		&\ 1\ & -0.67\\ 
		& & 1
\end{bmatrix}
\end{align} 
shown with their statistical uncertainty (first error), spread under mass-anisotropy variation (second error)
and corresponding correlation matrix.
Similarly, for the $K$-matrix parameterisation $(d)$ we have,
\begin{align} 
	\label{eq:3bar_Kmatrixe_fit_params}
	\begin{split}
		m_0   = &\ (0.52337 \pm 0.00036 \pm 0.00037) \cdot a_t^{-1} \\ 
		g_0   = &\ (0.81 \pm 0.15 \pm 0.20) \cdot a_t^{-1}\\   
		\gamma_0 = &\ (4.34 \pm 2.61 \pm 4.13)\\ 	
	\end{split} \
        \quad
	\begin{bmatrix} 
		\ 1\ \ & 0.34 & 0.30\\ 
		&\ 1\ & 0.98\\ 
		& & 1
\end{bmatrix} \, .
\end{align}
These fits are shown in Fig.~\ref{fig:3bar_kcot}, with their central values (solid line), 
statistical uncertainty (inner band) and spread from mass-anisotropy variations (outer bands).
The central values of all other reasonable fits ($\chisq \leq 2$) are also shown in red.
The reasonable parameterisations are in good agreement and the uncertainties are relatively small in the energy region where the amplitudes are constrained.
At higher kinematics, where the data is less constrained, there is greater spread of the reasonable fits. 
To better constrain this higher energy region, one would need to include the effects of the \D\etas\ channel.

\begin{figure}
	\centering
	\minipage{0.95\textwidth}
	\includegraphics[width=\linewidth]{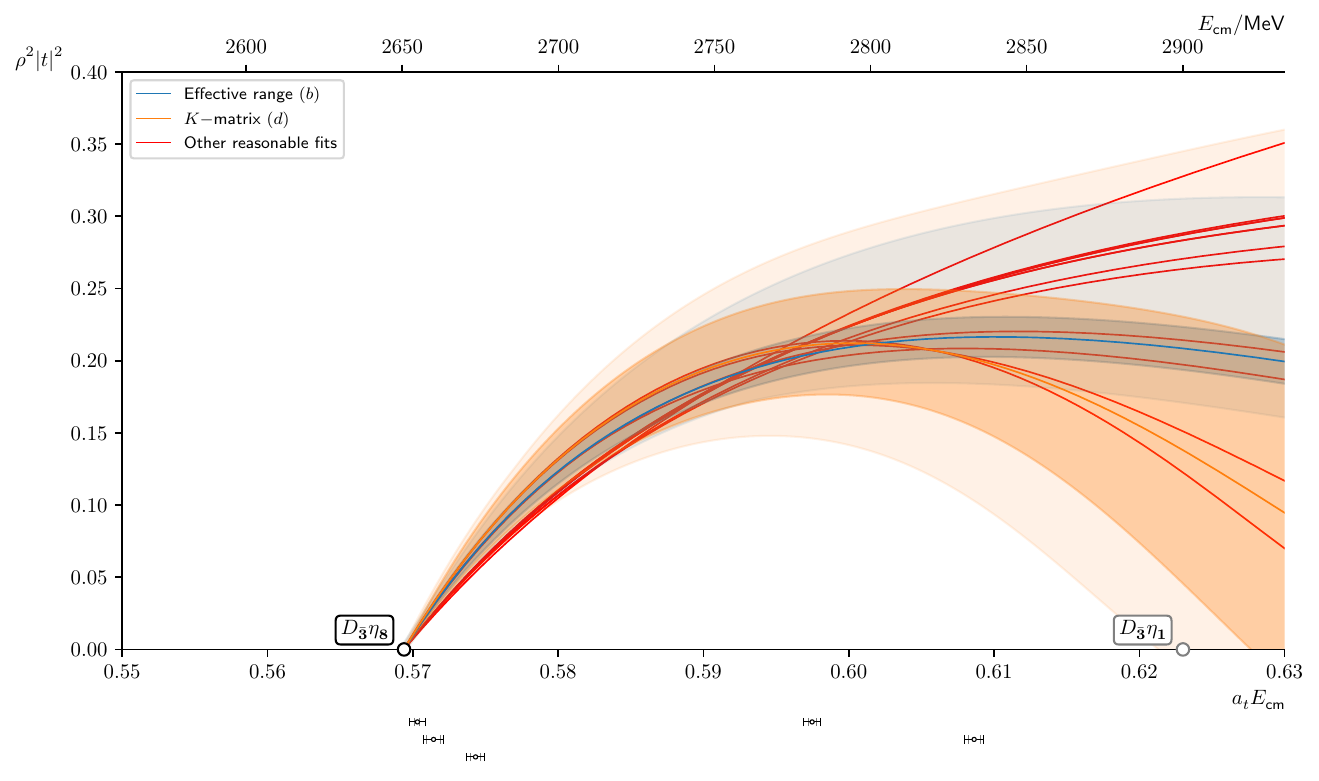}
	\caption{Amplitudes shown as $\rho^2|t|^2$ for elastic $S$-wave $\D\etaoct$ scattering in the flavour $\bar{\mathbf{3}}$ sector,
			 with curves and bands as in Fig.~\ref{fig:3bar_kcot}.
			 Points below the plot indicate energies at which the $t$-matrix has been constrained.
			 Circles on the horizontal axis indicate kinematic thresholds. 
     		}\label{fig:3bar_rhotsq}
    \endminipage
	\hfill
	\minipage{0.95\textwidth}
	\includegraphics[width=\linewidth]{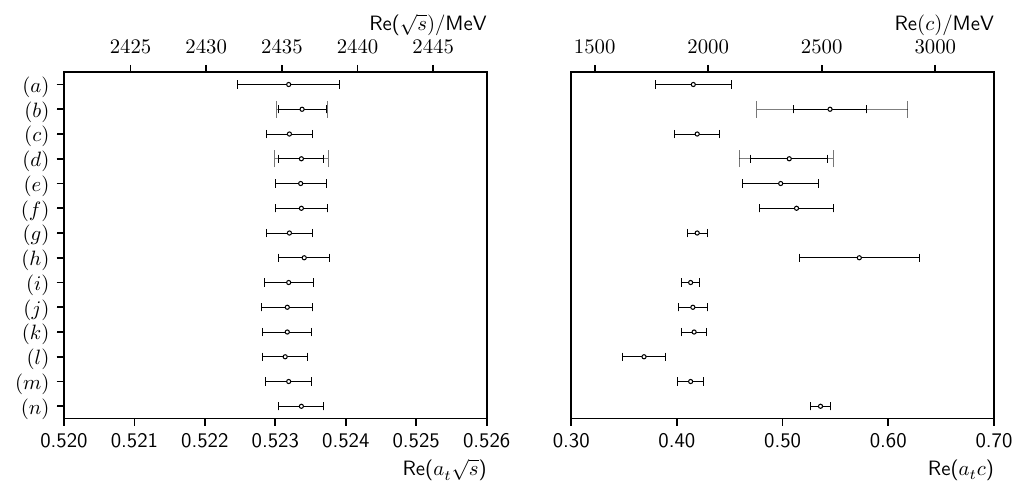}
	\caption{Physical-sheet pole positions (left) and associated coupling to the $\D\etaoct$ channel (right) in the flavour $\bar{\mathbf{3}}$ sector.
			 For each reasonable parameterisation, values are shown with their statistical error, 
			 and for the reference parameterisations an additional faded errorbar is presented 
			 corresponding to the spread of the values under mass-anisotropy variations.
			}\label{fig:3bar_poles}
	\endminipage
\end{figure}

The amplitudes are also presented as $\rho^2|t|^2$ in Fig.~\ref{fig:3bar_rhotsq}. 
Rapid turn on of the amplitude at scattering threshold is consistently seen across the reasonable fits and under mass-anisotropy variations. 
Such a characteristic in the amplitude is usually a sign of a pole nearby in the complex $s$-plane. 

The intersection of $k\cot\delta(k)$ with $-|k|$ for $k^2 <0$ in Fig.~\ref{fig:3bar_kcot}
signals the presence of a pole on the real axis below threshold on the physical sheet, 
i.e.\ a bound-state pole.
When analytically continuing the amplitude we indeed find a bound-state pole for each reasonable parameterisation,
with pole positions given in Table~\ref{table:3bar_fits} and plotted, alongside their corresponding couplings, in Fig.~\ref{fig:3bar_poles}.
The presence of a pole is a parameterisation-independent feature of the $t$-matrix and 
its location is consistent across different parameterisations and under mass-anisotropy variations.
However, the corresponding coupling is not as consistent across reasonable fits.
This is likely due to less constrained behaviour of the amplitudes between the bound state and threshold, 
making interpolation of the amplitude more susceptible to parameterisation dependence.
Taking an envelope over the minimal and maximum values from individual fits, we quote a range for the mass and coupling as

\begin{align*}
	a_tm &= 0.5225 - 0.5239 \text{ and } a_tc= 0.349 - 0.630
	\end{align*}
	or in physical units
	\begin{align*}
	m &= 2432 - 2439\ \text{MeV}\text{ and } c = 1620 - 2930\ \text{MeV}
	\end{align*}
No other poles were found in the energy region where the amplitudes were constained.\footnote{Some fits have poles on the unphysical sheet above inelastic threshold
 -- see supplementary material.}

\subsection{Flavour \texorpdfstring{$\mathbf{6}$}{Lg} Sector}
\label{section:6f}
\subsubsection{Finite-volume Spectra}
\label{section:6fspectra}

\begin{table}
	\centering
	\resizebox{0.75\columnwidth}{!}{%
		\begin{tabular}{|cccl|}
			\hline
			\rule{0pt}{12pt} $[000]\Lambda^P$ 	& $J^P$  & Channel  	& \SLJ{2S+1}{\ell}{J} \\
			\hline 
			\hline
			\rule{0pt}{12pt} $[000]A_1^+$ 	    & $ 0^+,\ \grey{ 4^+},...$	    & \D\etaoct  		& \SLJ{1}{S}{0} \\
										        &							    & \Dst\omoct  		& \grey{ \SLJ{1}{S}{0}, \SLJ{5}{D}{0},\ \SLJ{5}{D}{4}} \\[3pt]
											    &							    & \vdots 	        & \vdots  \\[3pt]
			\hline
			\rule{0pt}{12pt} $[000]T_1^-$ 	    & $ 1^-,\ \grey{3^-,\ 4^-,...}$	& \D\etaoct  		& \SLJ{1}{P}{1}, \grey{ \SLJ{1}{F}{3} } \\[3pt]
			   								    &								& \Dst\etaoct  	& \SLJ{3}{P}{1}, \grey{   \SLJ{3}{F}{3}, \SLJ{3}{F}{4} } \\[3pt]
											    &								& \D\omoct  		& \grey{ \SLJ{3}{P}{1}, \ \SLJ{3}{F}{3}, \SLJ{3}{F}{4} } \\[3pt]
											    &								& \vdots 	        & \vdots  \\[3pt]
			\hline
			\rule{0pt}{12pt} $[000]E^+$ 	    & $2^+,\ \grey{4^+,...}$		& \D\etaoct  	    & \SLJ{1}{D}{2} \\[3pt]
										        &						        & \Dst\etaoct  	& \SLJ{3}{D}{2} \\[3pt]
											    &							    & \D\omoct  		& \grey{ \SLJ{3}{D}{2} } \\[3pt]
											    &							    & \vdots 	        & \vdots  \\[3pt]
			\hline
			\rule{0pt}{12pt} $[000]T_2^+$ 	    & $2^+,\ \grey{3^+,\ 4^+,...}$	& \D\etaoct  		& \SLJ{1}{D}{2} \\[3pt]
										        &							    & \Dst\etaoct  	& \SLJ{3}{D}{2}, \SLJ{3}{D}{3}  \\[3pt]
											    &							    & \D\omoct  		& \grey{\SLJ{3}{D}{2}, \SLJ{3}{D}{3}} \\[3pt]
											    &							    & \vdots 	        & \vdots  \\[3pt]
			\hline
			\multicolumn{4}{c}{}  \\

			\hline
			\rule{0pt}{12pt} $[nmp]\Lambda$ 	& $|\lambda|^{(\tilde{\eta})}$   & Channel      	& \SLJ{2S+1}{\ell}{J} \\
			\hline 
			\hline
			\rule{0pt}{12pt} $[100]A_1$ 	 	& $0^+,\ \grey{4,...}$	   		 & \D\etaoct  		& \SLJ{1}{S}{0}, \SLJ{1}{P}{1}, \SLJ{1}{D}{2}, \grey{\SLJ{1}{F}{3}}\\[3pt]
												&				    	    	 & \Dst\etaoct  	& \SLJ{3}{P}{1}, \SLJ{3}{D}{2}, \grey{\SLJ{3}{F}{3}, \SLJ{3}{F}{4} } \\[3pt]
												&						         & \D\omoct  		& \grey{ \SLJ{3}{P}{1}, \SLJ{3}{D}{2}, \SLJ{3}{F}{3}, \SLJ{3}{F}{4} } \\[3pt]
												&						         & \vdots 	        & \vdots  \\[3pt]
			\hline
			\rule{0pt}{12pt} $[110]A_1$ 	    & $0^+,\ \grey{2,\ 4, ...}$ 	 & \D\etaoct  		        & \SLJ{1}{S}{0}, \SLJ{1}{P}{1}, \SLJ{1}{D}{2}, \grey{ \SLJ{1}{F}{3} }\\[3pt]
												&							     & \Dst\etaoct $^\dagger$  	& \SLJ{3}{P}{1}, \SLJ{3}{D}{2}, \SLJ{3}{D}{3}, \grey{\SLJ{3}{F}{3},  \SLJ{3}{P}{2}, \SLJ{3}{F}{2},  \SLJ{3}{F}{4} } \\[3pt]
												&							     & \D\omoct  	         	& \grey{ \SLJ{3}{P}{1}, \SLJ{3}{D}{2}, \SLJ{3}{F}{3},  \SLJ{3}{P}{2}, \SLJ{3}{F}{2}, \SLJ{3}{D}{3}, \SLJ{3}{F}{4} } \\[3pt]
											    &							     & \vdots 	                & \vdots  \\[3pt]
			\hline
	
		\end{tabular}
	    }
		\caption{As Table~\ref{table:3bar_subductions} but for the flavour $\mathbf{6}$ sector, 
				 including additional at-rest and in-flight finite-volume irreps.
				 Note that for the in-flight irreps, $\tilde{\eta}= P(-1)^J$ is only a good quantum number for zero helicity. $^\dagger$Since the $[110]A_1$ finite-volume energies used in the scattering analysis appear below \Dst\etaoct\ threshold, this irrep does not provide constraints on the \Dst\etaoct\ partial waves.
				 }\label{table:6f_subductions}
\end{table}

Next we discuss the flavour $\mathbf{6}$ sector 
which only has contributions from $D^{(*)}$-octet scattering. 
Table~\ref{table:6f_subductions} gives the infinite-volume scattering channel
contributions to various finite-volume irreps
with $G$-wave or higher and $J>4$ contributions not presented.
Again, staying below the first inelastic threshold of \Dst\omoct,
$[000]A_1^+$ contains information on elastic \D\etaoct\ scattering.

\begin{figure}
	\minipage{1\textwidth}
	\centering
	\includegraphics[width=0.64\linewidth]{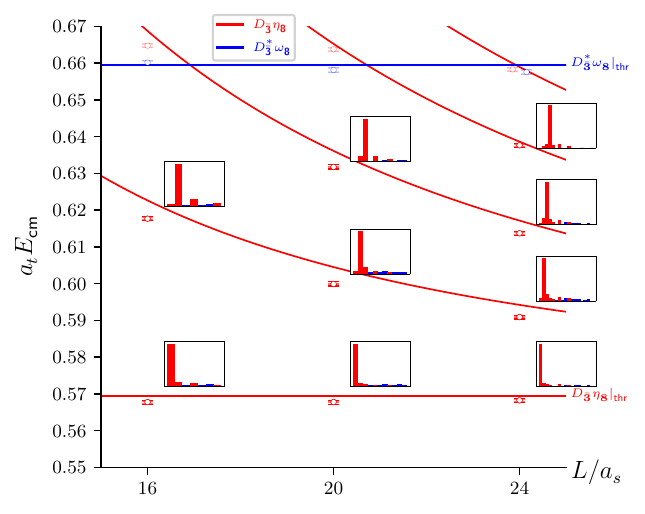}
	\caption{As Fig.~\ref{fig:3bar_spectra} but for the $[000]A^+_1$ irrep in the flavour $\mathbf{6}$ sector. Histograms show normalised operator overlaps with the meson-meson operators ordered by increasing non-interacting meson-meson energy.}
	\label{fig:6f_A1p_spectra}
  \endminipage
  \hfill
  \minipage{1\textwidth}
  \centering
  \includegraphics[width=0.85\linewidth]{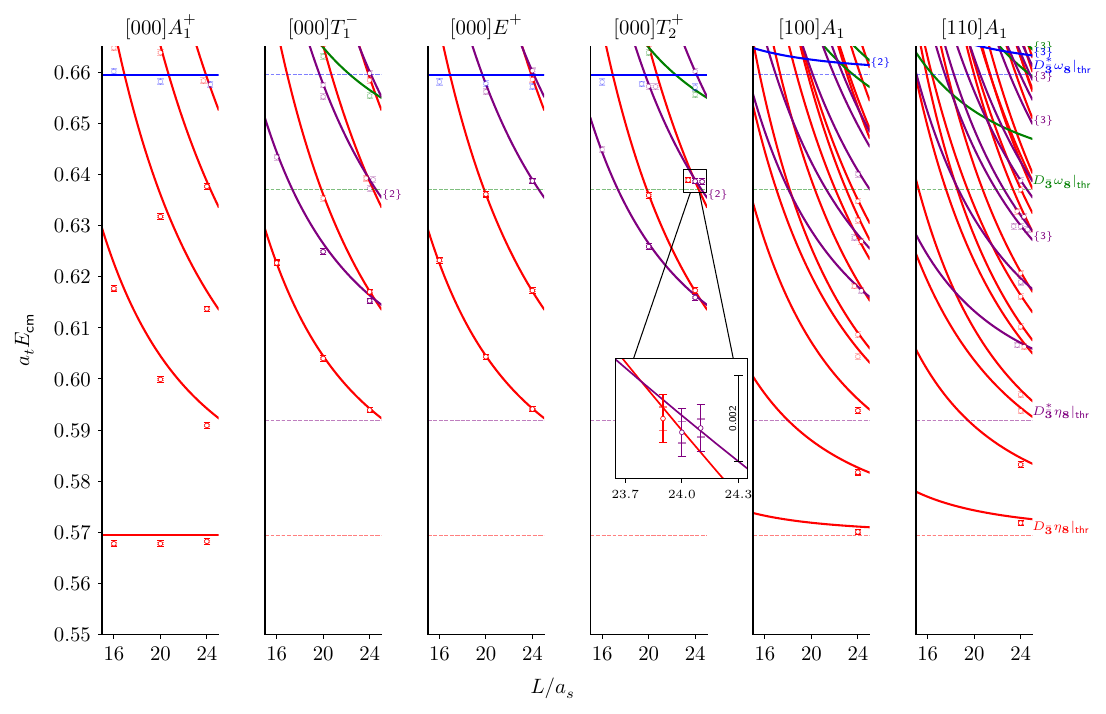}
  \caption{As Fig.~\ref{fig:3bar_spectra} but for each of 
		   the $[000]A^+_1$, $[000]T_1^-$, $[000]E^+$, $[000]T_2^+$, $[100]A_1$ and $[110]A_1$ irreps in the flavour $\mathbf{6}$ sector.
		   Colour coding is given by the threshold labels on the right hand side.
		   Thresholds that don't contribute in $S$-wave are indicated by dashed lines.
		   Degeneracies are labelled in $\{...\}$ when more than one,
		   with the insert showing a zoomed-in example in $[000]T_2^+$ irrep. 
  }\label{fig:6f_spectra}
  \endminipage
\end{figure}

After preliminary analysis using only the $[000]A_1^+$ irrep,
it was clear more data was needed to further constrain the $t$-matrix around \D\etaoct\ threshold,
and so data from the ``in-flight'' $[100]A_1$ and $[110]A_1$ irreps on the $L/a_s = 24$ lattice were included.
With further reduction of symmetry in these non-zero momentum irreps, 
additional partial waves contribute.
However, the lowest-lying additional partial waves, \D\etaoct(\SLJ{1}{P}{1}) and \D\etaoct(\SLJ{1}{D}{2}),\footnote{With higher partial waves assumed to be negligible.}
enter at $P$ and $D$ wave respectively
and so are expected to be suppressed at low energies in comparison to the $S$-wave contribution. 
To constrain these `background wave' channels,
we computed energy levels in the $[000]T_1^-$, $[000]E^+$ and $[000]T_2^+$ irreps
where they are the leading-order contributions.

This flavour sector has manifestly exotic flavour quantum numbers,\footnote{Meaning no fermion-bilinear operators 
can be constructed carrying these flavour quantum numbers.}
leaving only meson-meson operators to be used in the matrix of correlation functions.
Lists of operators are given in Tables \ref{table:6f_rest_list} and \ref{table:6f_in_flight_list} in Appendix \ref{appendix:interpolating_list}.
Fig.~\ref{fig:6f_A1p_spectra} shows the $[000]A_1^+$ spectrum, again with normalised operator overlaps. 
There are 9 levels well below the \Dst\omoct\ threshold with all but the $4^{\textrm{th}}$ level on the $24^3$ volume shifted downwards relative to non-interacting curves, suggesting attractive physics in this sector.
All 9 levels were used in the Lüscher analysis.
The spectra in all the irreps used for the $\mathbf{6}$ analysis are presented in Fig.~\ref{fig:6f_spectra}.
Again, numerical values for the energies are given in the supplementary material, and 
levels not used in the scattering analysis are faded out in the plots.

Moving to non-zero momentum, 
the $[100]A_1$ and $[110]A_1$ irreps have contributions from two additional channels below $\Dst\omoct$ threshold,
\Dst\etaoct\ and \D\omoct, which do not appear in the $[000]A_1^+$ irrep.
However, these enter in a minimum of $P$-wave and therefore have suppressed threshold behaviour,  
and their first non-interacting levels appear somewhat above the corresponding thresholds.
Only the lowest 3 energy levels in $[100]A_1$ and the lowest 2 in $[110]A_1$ are used in the scattering analysis. 
Similar to $[000]A_1^+$, downwards shifts compared to the non-interacting energy curves are observed for these levels.

The spectra from the 3 additional at-rest irreps are used to constrain the `background' partial waves. 
All levels appear very close to the non-interacting curves, suggesting weak interactions in these irreps.
Note that all finite-volume levels in these irreps appear above the first inelastic threshold of \Dst\etaoct\ scattering
and there are relatively low-lying levels with dominant overlap onto \Dst\etaoct\ operators,
highlighting the need to include this channel in the analysis of these irreps.
With the aim of constraining these amplitudes close to threshold, 
in the $[000]T_1^-$ irrep we use the lowest 6 points up to $a_tE = 0.63$ to fit to the channels \D\etaoct(\SLJ{1}{P}{1}) and \Dst\etaoct(\SLJ{3}{P}{1}). 
For the  $[000]E^+$ and $[000]T_2^+$ irreps, 
constraining 3 channels, \D\etaoct(\SLJ{1}{D}{2}), \Dst\etaoct(\SLJ{3}{D}{2}) and \Dst\etaoct(\SLJ{3}{D}{3}),
we use a slightly higher cutoff of $a_tE = 0.64$ resulting in 6 levels from $[000]E^+$ and 7 levels from $[000]T_2^+$.

\subsubsection{\texorpdfstring{$\D\etaoct\ S$-wave scattering}{Lg}}
\label{section:6f_scalar_results}

As mentioned above, the low lying spectra of the in-flight irreps are expected to be dominated by the $S$-wave \D\etaoct\ amplitude.
As such, for now we neglect background partial wave contributions to the in-flight irreps and consider a single \D\etaoct(\SLJ{1}{S}{0}) channel
constrained by 9 levels from $[000]A_1^+$, 3 levels from $[100]A_1$ and 2 levels from $[110]A_1$.
We will test this approximation and show that it is justified in the next section.

\begin{figure}
	\minipage{\textwidth}
	\includegraphics[width=0.95\linewidth]{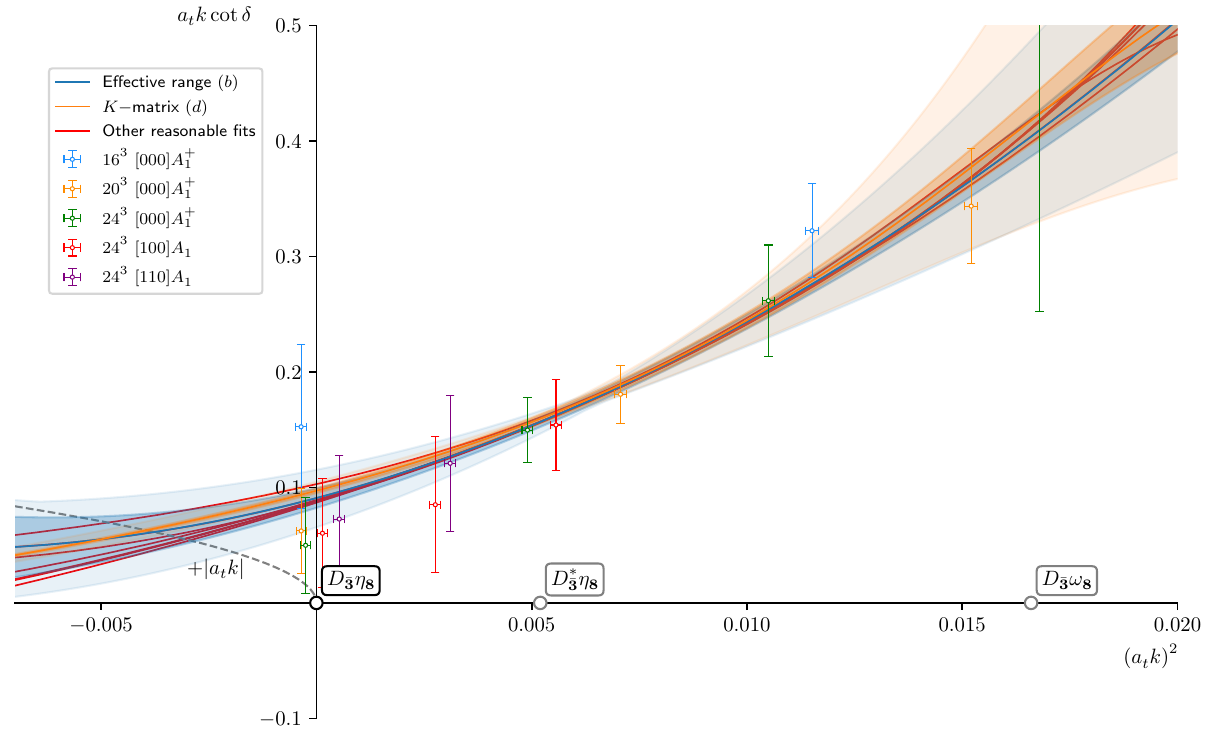}
	\caption{As Fig.~\ref{fig:3bar_kcot} but for the flavour $\mathbf{6}$ sector.
			 Additional grey thresholds indicate contributions to the in-flight irreps 
			 which don't enter $[000]A_1^+$.
			 }\label{fig:6f_scalar_kcot}
	\endminipage
\end{figure}

\begin{table}
	\centering
	\resizebox{0.85\columnwidth}{!}{%
	\centering
			\begin{tabular}{lcc}
				\hline
				\multicolumn{1}{c}{\rule{0pt}{11.5pt}Parameterisation}  & \multicolumn{1}{c}{$\chi^2 / N_{\rm{dof}} \ $ } & \multicolumn{1}{c}{$a_t\sqrt{s_{\rm{pole}}}$} \\
				\hline
				Effective range                                             										 &                                     &                                            \\[5pt] 
				$(a)\ k\cot\delta(k) = \frac{1}{a} + \frac{1}{2}rk^2  $ 				  				             & \grey{$\frac{29.03}{14-2}= 2.42$}   &                                            \\[8pt]

				$\bm{(b)\ k} \textbf{ cot} \bm{\delta(k) = \frac{1}{a} + \frac{1}{2}rk^2  + P_2k^4}$ 				 & $\bm{\frac{13.47}{14-3}= 1.22}$      & $\bm{0.55155(748)}$                       \\[8pt]
				\hline
				$K$-matrix with Chew-Mandelstam phase-space                                                          &                               	   &                                            \\[5pt] 
				$(c)\ K(s) = \gamma_0 + \gamma_1 s $    		                   								 & \grey{$\frac{31.10}{14-2}= 2.59$}   &  			                                \\[8pt]

				$\bm{(d)\ K(s) = \gamma_0 + \gamma_1 s + \gamma_2 s^2}$         			             & $\bm{\frac{13.91}{14-3} = 1.26}$ 	   & $\bm{0.55066(43)}$                         \\[8pt]

				$(e)\ K^{-1}(s) = c_0 + c_1 s + c_2 s^2	$              				          		     & $\frac{13.50}{14-3} = 1.23$          & 0.55338(169)                               \\[8pt]

				$(f)\ K^{-1}(s) = \frac{c_0+ c_1s}{1+d_1s}$                 				      		 & $\frac{12.49}{14-3} = 1.14$          & 0.55561(105)                               \\[8pt]

				$(g)\ K^{-1}(s) = \frac{c_1s+ c_2s^2}{1+d_1s}$                			         & $\frac{12.61}{14-3} = 1.15$          & 0.55494(98)                               \\[8pt]

				\hline
				$K$-matrix with simple phase-space                                                                   &                               	   &                                            \\[5pt] 

				$(h)\ K(s) = \gamma_0 + \gamma_1 s$            					           		             & \grey{$\frac{41.15}{14-2}= 3.43$}   &  			                                \\[8pt]

				$(i)\ K(s) = \gamma_0 + \gamma_1 s + \gamma_2 s^2$                  			         & $\frac{15.07}{14-3} = 1.37$          & 0.54623(42)                                \\[8pt]
	
				$(j)\ K^{-1}(s) = c_0 + c_1 s + c_2 s^2 $                				        	   	 & $\frac{13.44}{14-3} = 1.22$          & 0.55159(201)                               \\[8pt]

				$(k)\ K^{-1}(s) = \frac{c_0+ c_1s}{1+d_1s} $                  				      		 & $\frac{12.60}{14-3} = 1.15$          & 0.55508(107)                              \\[8pt]

				$(l)\ K^{-1}(s) = \frac{c_1s +c_2s^2}{1+d_1s} $               			         & $\frac{12.76}{14-3} = 1.16$          & 0.55418(110)                              \\[8pt]

				\hline 

			\end{tabular}
	}
	\caption{As Table~\ref{table:3bar_fits} but for the flavour $\mathbf{6}$ sector and with virtual bound-state pole locations.
			 Reference parameterisations are in bold. 
			 Rejected fits with $\chi^2/ N_{\rm{dof}} > 2 $ are in grey with no pole position given. 
			 } \label{table:6f_fits} 
\end{table} 

As for the flavour $\bar{\mathbf{3}}$ sector, 
we obtain the corresponding phase-shift for each finite-volume energy and present these as $k \cot \delta(k)$ in Fig.~\ref{fig:6f_scalar_kcot}.
To parameterise the energy dependence of the scattering amplitude, we consider various effective range and $K$-matrix forms.
Due to the small but non-negligible curvature of the $k\cot\delta(k)$ data, only functions with at least 3 parameters gave reasonable \chisq $\ (\leq 2)$.
To increase the diversity of reasonable fits, $K$-matrix parameterisations featuring a ratio of polynominals were also considered. 
The parameterisations used are summarised in Table~\ref{table:6f_fits}.
Again, to illustrate the results, effective range $(b)$ and $K$-matrix $(d)$ are chosen as reference parameterisations.
The effective range $(b)$ fit parameters are, 
\begin{figure}
	\minipage{1\textwidth}
	\includegraphics[width=\linewidth]{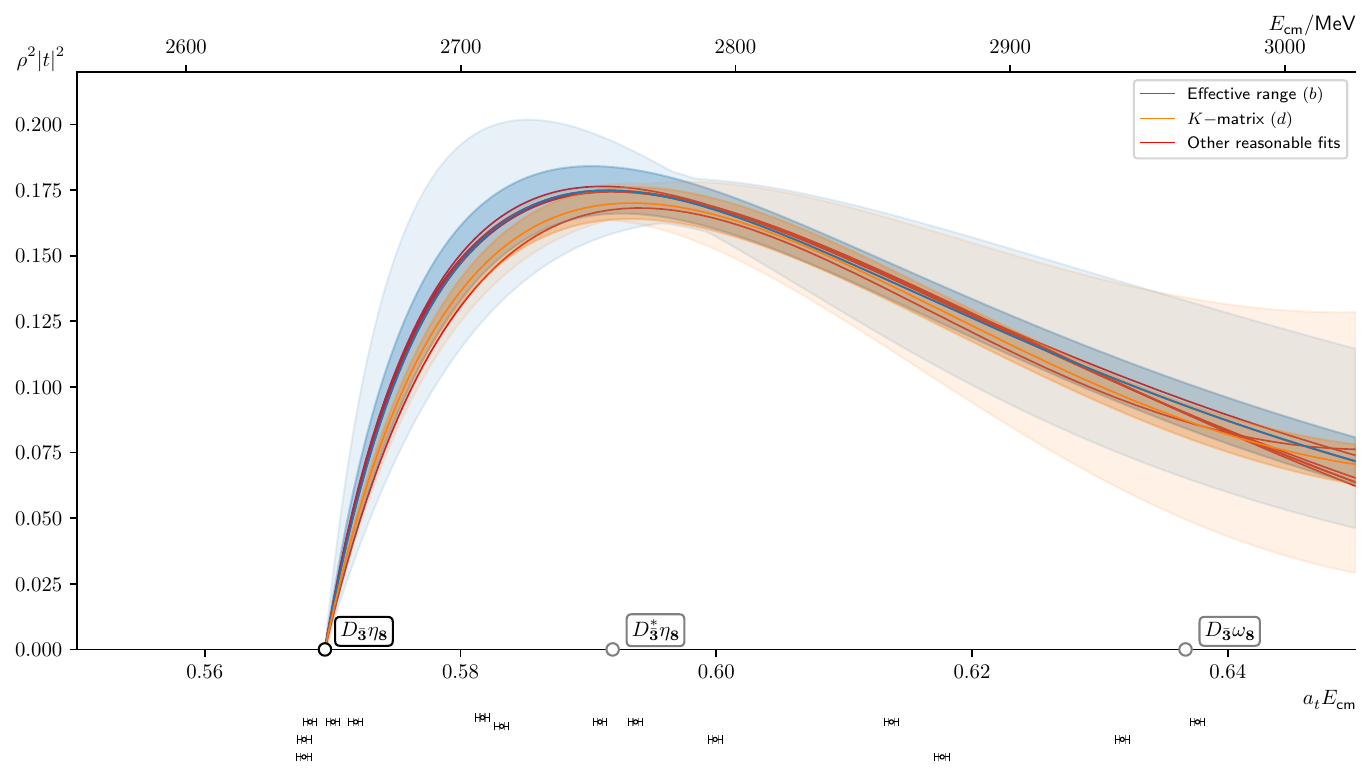}
	\caption{As Fig.~\ref{fig:3bar_rhotsq} but for the flavour $\mathbf{6}$ sector.
			 Grey thresholds indicate additional contributions to the in-flight irreps 
			 which don't enter in $[000]A_1^+$.
			 }\label{fig:6f_scalar_rhotsq}
	\endminipage
	\hfill
	\minipage{1\textwidth}
	\includegraphics[width=\linewidth]{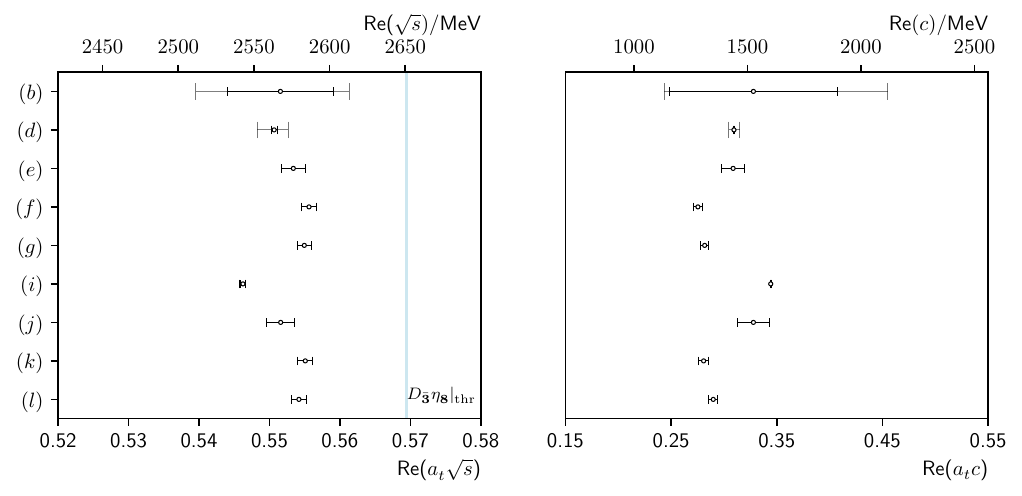}
	\caption{As Fig.~\ref{fig:3bar_poles} but showing the location and coupling of the virtual bound-state pole in the flavour $\mathbf{6}$ sector. 
			 }	\label{fig:6fpoles}
\endminipage
\end{figure}
\begin{align} 
	\label{eq:6f_effrange_fit_params}
	\begin{split}
		a   = &\ (11.02 \pm 1.32 \pm 3.91) \cdot a_t \\ 
		r   = &\ (19.70 \pm 4.46 \pm 6.08) \cdot a_t\\   
		P_2 = &\ (542 \pm 164 \pm 385) \cdot a^{3}_t\\ 	
	\end{split} \
        \quad
        \begin{bmatrix} 
		\ 1\ \ & -0.88 & -0.96\\ 
		&\ 1\ & 0.92\\ 
		& & 1
\end{bmatrix}
\end{align} 
with uncertainty notation as in Eq.~\ref{eq:3bar_effrange_fit_params}.
For the $K$-matrix $(d)$ the fit parameters are, 
\begin{align} 
	\label{eq:6f_Kmat_fit_params}
	\begin{split}
		\gamma_0  = &\ (34.21 \pm 0.22 \pm 5.43) \\ 
		\gamma_1  = &\ (-152.97 \pm 0.67 \pm 25.29) \cdot a^{2}_t\\   
		\gamma_2  = &\ (174.23 \pm 1.36 \pm 29.20) \cdot a^{4}_t\\ 	
	\end{split} \
        \quad
	\begin{bmatrix} 
	\ 1\ \ & -0.63 & -0.25\\ 
	&\ 1\ & -0.58\\ 
	& & 1
\end{bmatrix}\, .
\end{align} 
From these fit parameters it can be seen that the systematic uncertainties are more important than the statistical uncertainties in this sector.
Fig.~\ref{fig:6f_scalar_kcot} shows the amplitudes' central values, statistical uncertainty (inner band) and uncertainty from mass-anisotropy variations (outer band), as well as 
the central values of the other reasonable fits (red curves). 
Again, all reasonable fits are in good agreement, particularly around threshold where the in-flight data helps constrain the amplitude further.
Despite the reference parameterisations appearing to have small statistical uncertainties, they display larger spread under mass-anisotropy variations. 

Once more the amplitudes are displayed as $\rho^2|t|^2$ in Fig.~\ref{fig:6f_scalar_rhotsq}.
There is excellent agreement between the reasonable fits at constrained kinematics 
with all central values appearing inside the reference parameterisations' narrow statistical bands.
Although slightly smaller than in the flavour $\bar{\mathbf{3}}$ sector,
all reasonable fits demonstrate rapid growth at $\D\etaoct$ threshold, suggesting the presence of a nearby pole.

In Fig.~\ref{fig:6f_scalar_kcot} we note that the reasonable fits all intersect with $+|k|$ for $k^2<0$, 
indicating the presence of a virtual bound state.
Upon analytically continuing to the complex $s$-plane,
a virtual bound state was indeed consistently seen for each reasonable fit. 
The pole locations are summarised in Table~\ref{table:6f_fits}, 
and plotted with their corresponding couplings in Fig.~\ref{fig:6fpoles}.
The pole locations and couplings are precisely determined for each parameterisation, but there appears to be some tension between different parameterisations.
This may in part be due to not having data below threshold (unlike in the flavour $\bar{\mathbf{3}}$ sector). 
Furthermore, systematic uncertainty may be more significant here
as seen by the mass-anisotropy uncertainties being much larger than the statistical uncertainty.
As a conservative estimate for the value for the pole location and its coupling to \D\etaoct, 
the values from the effective range fit and its uncertainty are chosen, which encomposes the values from all the other reasonable fits. 
This gives, 

\begin{align*}
	a_t\sqrt{s_{\rm{pole}}} &= 0.5396 - 0.5614 \text{ and } a_tc = 0.243- 0.455
\end{align*}
or in physical units, 
\begin{align*}
	\sqrt{s_{\rm{pole}}} &= 2510 -  2610\ \text{MeV} \text{ and } c = 1130 - 2120\ \text{MeV}
\end{align*}
No other poles were found in the energy region constrained.

\subsubsection{Effect of background partial waves}

We now investigate the potential impact of the other partial waves that can contribute to the non-zero momentum irreps and test the assumption that the \D\etaoct(\SLJ{1}{S}{0}) channel dominates.
We start by determining the background $J^P =1^-$ amplitudes. 
Considering a $t$-matrix with two channels, \D\etaoct(\SLJ{1}{P}{1}) and \Dst\etaoct(\SLJ{3}{P}{1}),
constrained by 6 finite-volume energies in the $[000]T_1^-$ irrep, 
it was found that a $K$-matrix parameterisation with Chew-Mandelstam phase-space and constant components 
could describe the data well, resulting in a minimised $\chisq= \frac{1.21}{6-3} = 0.40$ with fit parameters
\begin{equation}
	\label{eq:vector_decple_parameters}
	\begin{aligned}
		\gamma_0\{\D\etaoct(\SLJ{1}{P}{1})| \D\etaoct(\SLJ{1}{P}{1})\}	 & = (0.92 \pm 1.70)  \\ 
		\gamma_0\{\D\etaoct(\SLJ{1}{P}{1})| \Dst\etaoct(\SLJ{3}{P}{1})\}	 & = (8.57 \pm 2.41)  \\ 
		\gamma_0\{\Dst\etaoct(\SLJ{3}{P}{1})| \Dst\etaoct(\SLJ{3}{P}{1})\} & = (14.41 \pm 3.36)  \\
	\end{aligned} \
        \quad
        \begin{bmatrix} 
		& 1& 0.86 & 0.09\\ 
		& &1 & -0.04 \\
		& & & 1
	\end{bmatrix}\, .
\end{equation}
A $K-$matrix with off-diagonal components fixed to zero also described the data well with a \chisq $=\frac{4.14}{6-2} = 1.04$.

To obtain constraints on the $J^P = 2^+$, $3^+$ contributions, 
we consider a block diagonal 3-channel $t$-matrix, \D\etaoct(\SLJ{1}{D}{2}), \Dst\etaoct(\SLJ{3}{D}{2}) and \Dst\etaoct(\SLJ{3}{D}{3}), 
and fit to 13 energy levels in the $[000]E^+$ and $[000]T_2^+$ irreps.
It was found that a decoupled constant $K$-matrix parameterisation with Chew-Mandelstam phase-space again described the finite-volume spectra well, 
with fit parameters, 
\begin{equation}
	\label{eq:tensor_decple_parameters}
	\begin{aligned}
		\gamma_0\{ \D\etaoct(\SLJ{1}{D}{2})| \D\etaoct(\SLJ{1}{D}{2})\}	 & = (-1.06 \pm 7.28)  \\ 
		\gamma_0\{ \Dst\etaoct(\SLJ{3}{D}{2})| \Dst\etaoct(\SLJ{3}{D}{2})\} & = (45.7 \pm 29.7)  \\
		\gamma_0\{  \Dst\etaoct(\SLJ{3}{D}{3})| \Dst\etaoct(\SLJ{3}{D}{3})\}	 & = (44.6 \pm 26.1)  \\ 
	\end{aligned} \
        \quad
        \begin{bmatrix} 
		\ 1\ \ & 0.58 & 0.06\\ 
		&\ 1\ & -0.31\\ 
		& & 1
	\end{bmatrix}
\end{equation}
and $\chisq = \frac{9.09}{13-3} = 0.91$.
A $K$-matrix with a non-fixed off-diagonal entry for the $J^P =2^+$ sector was also considered,  
but this entry was found to be consistent with zero,
$\gamma_0\{ \Dst\etaoct(\SLJ{3}{D}{2})| \D\etaoct(\SLJ{1}{D}{2})\}	 = (-1.0\times 10^{-3} \pm 12)$, with $\chisq=\frac{9.09}{13-4} = 1.01$.

\begin{figure}
	\minipage{1\textwidth}
	\includegraphics[width=\linewidth]{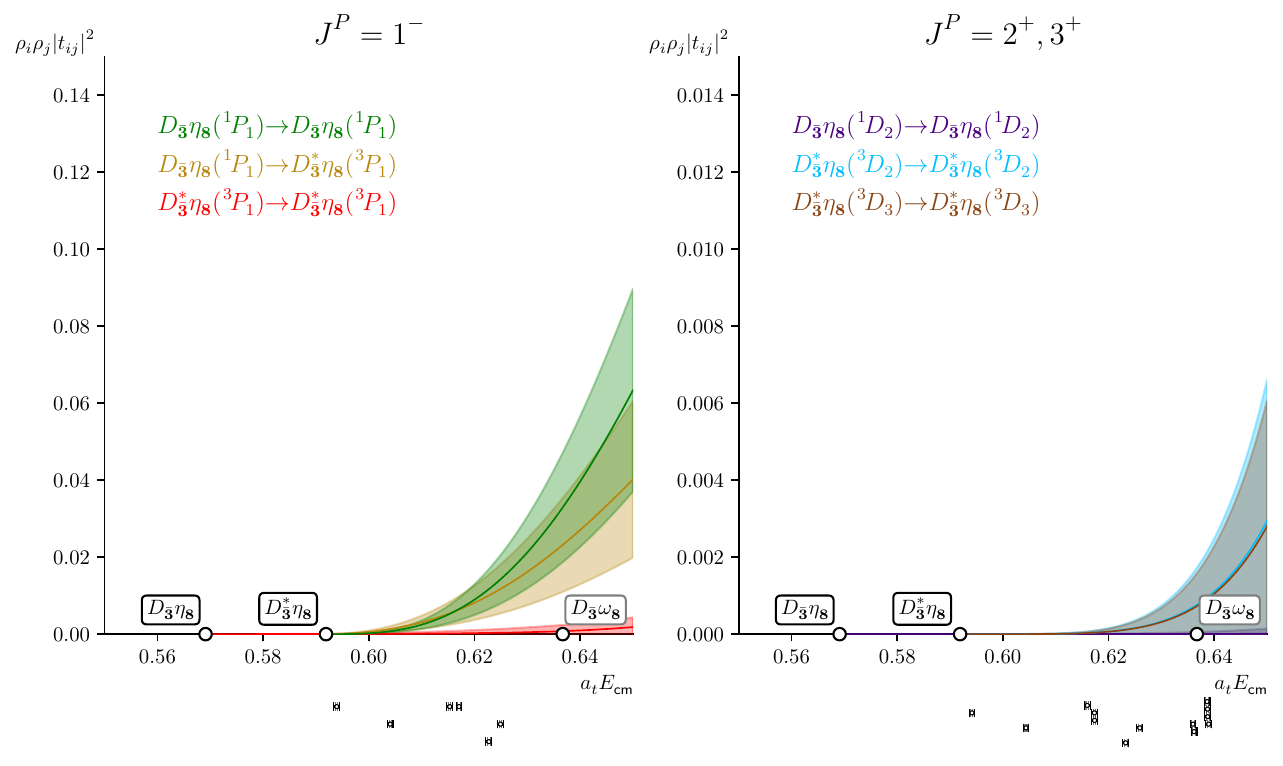}
	\caption{As Fig.~\ref{fig:6f_scalar_rhotsq} but for the $J^P =1^-$ channels (left) and $J^P =2^+,3^+$ channels (right), 
			 each for a single $K$-matrix parameterisation, 
			 with bands corresponding to statistical uncertainties.
			 }\label{fig:6f_background}
	\endminipage
\end{figure}

The $J^P = 1^-$ and $2^+$, $3^+$ amplitudes, i.e. Eq.~\ref{eq:vector_decple_parameters} and Eq.~\ref{eq:tensor_decple_parameters}, are plotted in Fig.~\ref{fig:6f_background}.
As expected, all amplitudes are very small near threshold with the $D$-wave amplitudes displaying more suppression than the $P$-wave amplitudes.
Importantly, at low energies ($a_tE_{\sf cm}<0.62$) these amplitudes are much smaller than the corresponding $S$-wave \D\etaoct\ amplitude.
Since the in-flight levels used for the $J^P=0^+$ analysis appear below $a_tE =0.60$, 
this supports the validity of the approximation that the background waves are not important when fitting to those in-flight levels.

\begin{figure}
	\minipage{1\textwidth}
	\centering
	\includegraphics[width=0.95\linewidth]{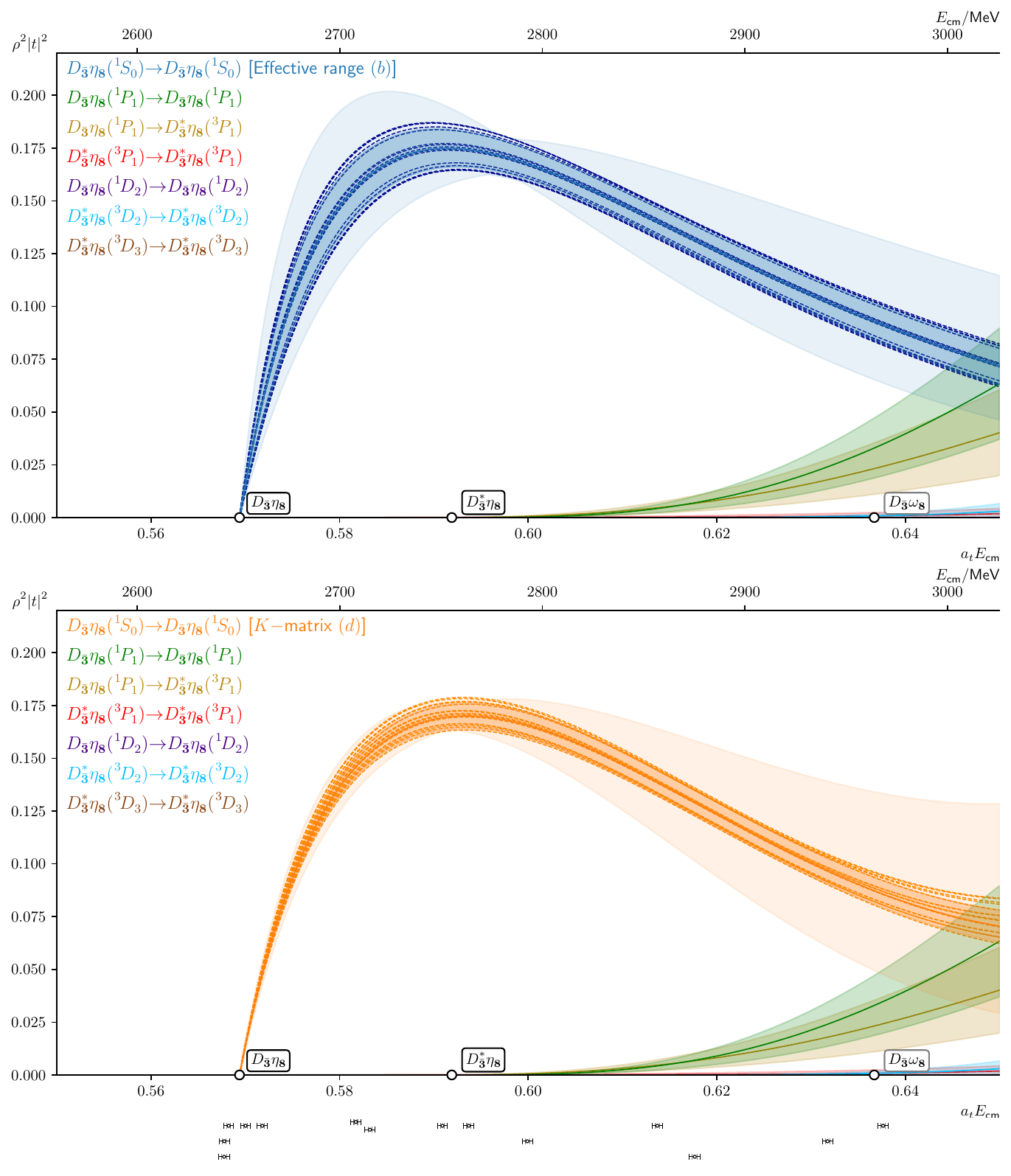}
	\caption{Comparison of $J^P=0^+$ amplitudes from effective range $(b)$ (top) and $K$-matrix $(d)$ (bottom) parameterisations, with and without the inclusion of fixed non-zero background waves 
			as described in the text.
			Background waves are also plotted to compare size of amplitudes.
			 }\label{fig:6f_rhotsq_compare}
	\endminipage
\end{figure}

With the background waves constrained, we repeated the $J^P=0^+$ analysis of Section~\ref{section:6f_scalar_results} for the reference parameterisations using fixed non-zero background partial waves.
Along with the $1^-$, $2^+$ and $3^+$ amplitudes with parameters fixed to the central values from Eq.~\ref{eq:vector_decple_parameters} and Eq.~\ref{eq:tensor_decple_parameters}, we also considered amplitudes where those parameters were varied by $1\sigma$ resulting in 5 fits.\footnote{To be more precise, 
given the determined fit parameters of the $J^P=1^-$ amplitudes, $\alpha^*\pm \sigma_{\alpha}$, and fit parameters of $J^P=2^+,3^+$ amplitudes, $\beta^*\pm \sigma_{\beta}$,
the four combinations of $\pm1\sigma$ on each of the sets of parameters are considered as  fixed inputs for the $J^P = 0^+$ analysis.}
The fit parameters for $J^P = 0^+$ effective range parameterisation $(b)$
with fixed background partial waves are
\begin{align} 
	\label{eq:Eff_range_bg}
\begin{split}
	a^{(bg)}   =\ & (11.09 \pm 1.34 \pm 1.54) \cdot a_t \\ 
	r^{(bg)}   =\ & (19.87 \pm 4.46 \pm 4.93) \cdot a_t\\   
	P^{(bg)}_2 =\ &  (536 \pm 164 \pm 181) \cdot a^{3}_t\\ 	
\end{split}\
\quad
\begin{bmatrix} 
	\ 1\ \ & -0.88 & -0.96\\ 
	&\ 1\ & 0.92\\ 
	& & 1
\end{bmatrix}
\end{align} 
where the central value and first uncertainty come from the fit using central values from Eq.~\ref{eq:vector_decple_parameters} and Eq.~\ref{eq:tensor_decple_parameters} as the fixed background wave parameters,
and the second uncertainty comes from an envelope over the central value and uncertainty when considering $1\sigma$ variations of these background wave fit parameters.
The range of $\chi^2$ values for these fits is $\chisq  = 1.18 - 1.24$.
It can be seen that the fit parameters for the effective range parameterisation $(b)$ are equal within statistical uncertainties to those in Section~\ref{section:6f_scalar_results} where the background waves were set to zero -- the presence of background partial waves induces hardly any changes to the fit parameters.
For the $K$-matrix parameterisation $(d)$ we have 
\begin{align} 
	\label{eq:Kmat_bg}
\begin{split} 
	\gamma^{(bg)}_0  = &\ (35.40 \pm 0.22 \pm 1.60) \\ 
	\gamma^{(bg)}_1  = &\ (-159.19 \pm 0.66 \pm 7.87) \cdot a^{2}_t\\   
	\gamma^{(bg)}_2  = &\ (182.41 \pm 1.35 \pm 10.77) \cdot a^{4}_t\\ 	
\end{split} \
\quad
\begin{bmatrix} 
	\ 1\ \ & -0.63 & -0.27\\ 
	&\ 1\ & -0.57\\ 
	& & 1
\end{bmatrix}
\end{align} 
with corresponding $\chisq  = 1.24 - 1.27$.
Comparing these $K-$matrix parameterisation $(d)$ fit parameters to the corresponding ones in Section~\ref{section:6f_scalar_results} we see small shifts but the values are consistent within the uncertainty from mass-anisotropy variations.

This can also be seen in Fig.~\ref{fig:6f_rhotsq_compare} where, for the effective range $(b)$ (top panel) and $K$-matrix $(d)$ (bottom panel) parameterisations, fits with and without fixed background wave contributions are presented.
On each plot, the resulting amplitudes when non-zero fixed background partial waves were used are shown as dashed lines: for each of the 5 fits described above the middle dashed line shows the central value of the $0^+$ amplitude and the upper and lower dashed lines show the statistical uncertainty on this amplitude.
The $0^+$ amplitudes from fits where background partial waves are set to zero are shown as coloured bands in the same way as in Fig.~\ref{fig:6f_scalar_rhotsq}.
The fixed $J^P =1^-, 2^+, 3^+$ background wave amplitudes are also plotted with coloured bands showing their statistical uncertainties.
From the plot it can be seen that the fits of the $J^P =0^+$ sector vary by only a small amount if non-zero background partial wave amplitudes are allowed for. 
The mass-anisotropy variations in Section~\ref{section:6f_scalar_results} give a larger uncertainty on the results.

The corresponding $J^P =0^+$ virtual bound-state pole locations when using fixed non-zero background partial waves are $a_t\sqrt{s_{\rm{pole}}} = 0.5519 \pm 0.0074 \pm 0.0090$ for the effective range $(b)$ parameterisation and 
$a_t\sqrt{s_{\rm{pole}}} = 0.55096 \pm 0.00042 \pm 0.00082$ for the $K$-matrix $(d)$, 
with the same uncertainty notation as in Eqs.~\ref{eq:Eff_range_bg} and \ref{eq:Kmat_bg}.
We see that these values are equal to the corresponding values in Table~\ref{table:6f_fits} within statistical uncertainties.
We conclude that the results from the fits for the $0^+$ amplitudes in Section~\ref{section:6f_scalar_results} show very little background wave dependence in the energy region constrained 
and the approximation that the low-lying levels from the in-flight irreps are dominated by \D\etaoct(\SLJ{1}{S}{0}) is well validated.
Therefore, the results of our analysis using a single \D\etaoct\ scattering channel, i.e.\ finding a virtual bound-state in the flavour $\mathbf{6}$, are robust.

\subsection{Flavour \texorpdfstring{$\overline{\mathbf{15}}$}{Lg} Sector}
\subsubsection{Finite-volume spectrum}

\begin{figure}
	\centering
	\minipage{0.8\textwidth}
	  \includegraphics[width=\linewidth]{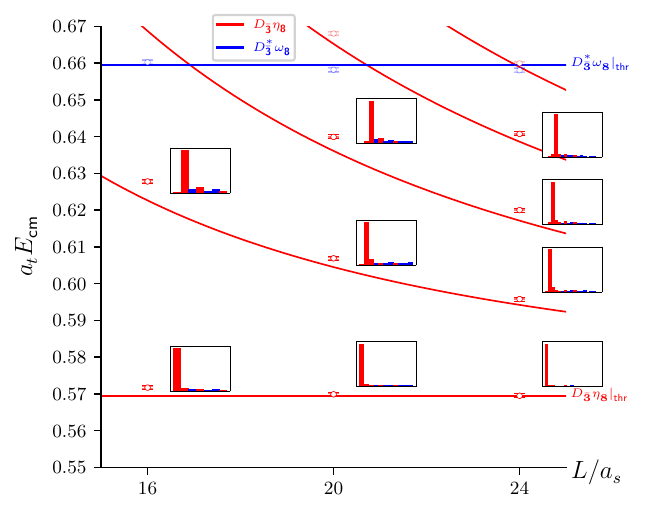}
	\endminipage
	\caption{As Fig.~\ref{fig:3bar_spectra} but for the $[000]A_1^+$ irrep in the flavour $\overline{\mathbf{15}}$ sector, with 
			histograms showing normalised overlaps with meson-meson operators corresponding to lower non-interacting energies.
			}\label{fig:15bar_spectra}
\end{figure}

We now discuss the other exotic flavour sector, the $\overline{\mathbf{15}}$.
The infinite-volume scattering channel contributions and meson-meson operators
are the same as for the $\mathbf{6}$ irreps,
and we refer back to Table~\ref{table:6f_subductions} and Table~\ref{table:6f_rest_list} in Appendix \ref{appendix:interpolating_list} for these.
Staying well below the first inelastic threshold, 
data from the $[000]A^+_1$ irrep is used to investigate $S$-wave \D\etaoct\ scattering.

\begin{figure}
	\minipage{1\textwidth}
	\centering
	\includegraphics[width=0.85\linewidth]{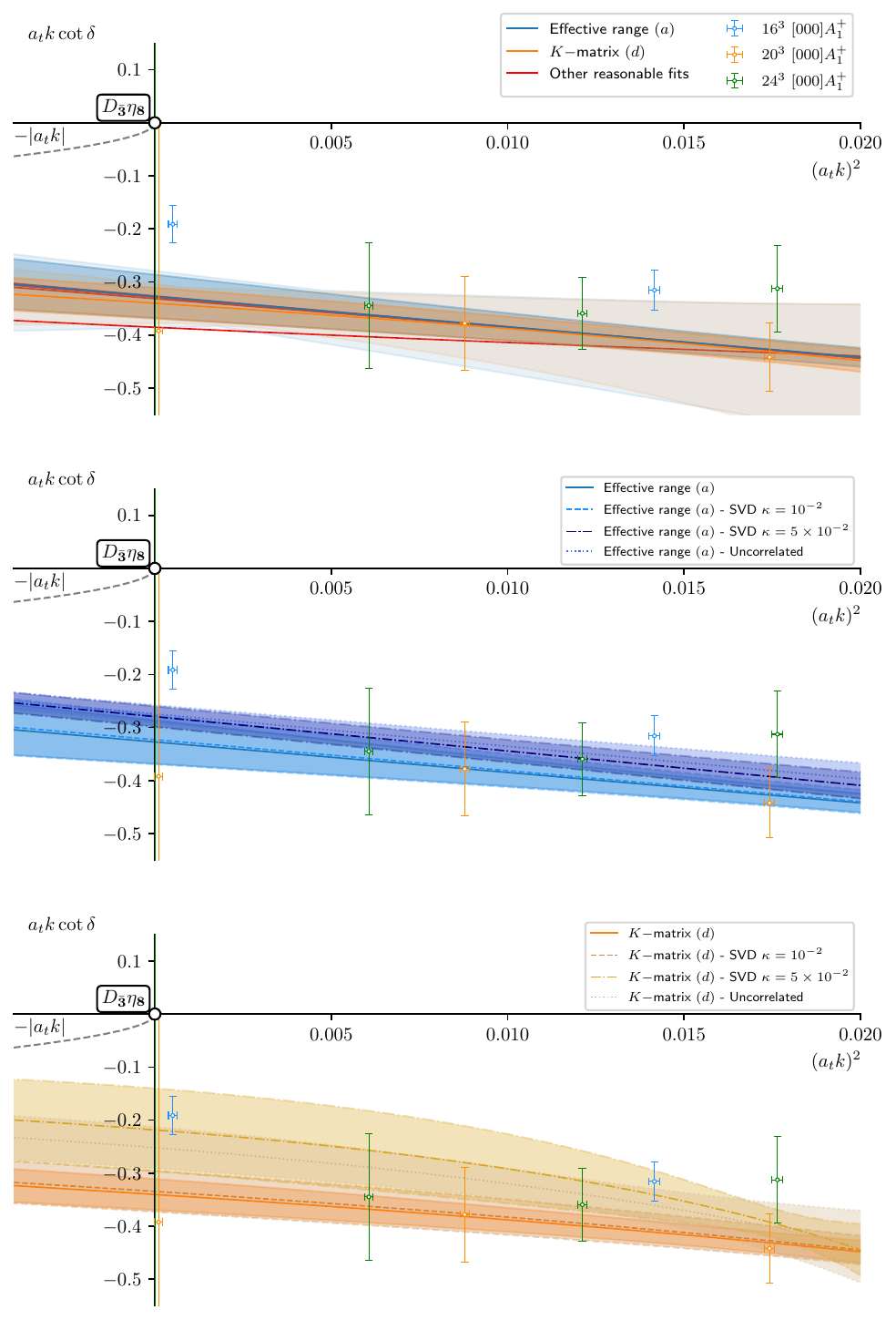}
	\caption{Upper: As Fig.~\ref{fig:3bar_kcot} but for the flavour $\overline{\mathbf{15}}$ sector.
			 Middle: Comparison of SVD cut and uncorrelated fits for the effective range parameterisation $(a)$. 
			 Lower: Comparison of SVD cut and uncorrelated fits for the $K$-matrix parameterisation $(d)$.
			 }\label{fig:15bar_kcot_all_fits_and_SVD_uncorr_fits}
	\endminipage
\end{figure}

The finite-volume spectrum is shown alongside non-interacting meson-meson curves in Fig.~\ref{fig:15bar_spectra},
again with the inclusion of histograms showing normalised overlaps of the lowest non-interacting energy meson-meson operators.
It can be seen that there is a one-to-one correspondence between computed energies and non-interacting energies,
with computed energies lying slightly above the non-interacting curves,
suggesting weak repulsive interactions.
Using energies well below the first inelastic threshold,
there are 9 levels that can be used to determine the scattering amplitude. 
Numerical values of the energies and associated uncertainties can be found in the supplementary material.

\subsubsection{Scattering amplitudes and poles}

As for the other two flavour sectors,
the phase shifts corresponding to each individual energy level are presented as $k\cot\delta(k)$ in Fig.~\ref{fig:15bar_kcot_all_fits_and_SVD_uncorr_fits}.
There appears to be no significant energy dependence and the points lie roughly on a straight line,
with rough extrapolation below threshold suggesting there are no poles in the energy region constrained.

Parameterisations fit to the data are summarised in Table~\ref{table:15bar_fits} and plotted in the upper panel of Fig.~\ref{fig:15bar_kcot_all_fits_and_SVD_uncorr_fits}.
Central values, statistical uncertainties and systematic uncertainties are presented in the figure for two reference parameterisations, effective range $(a)$ and $K$-matrix $(d)$,
as well as central values for other reasonable parameterisations.\footnote{
	Some parameterisations we use, labelled with $\ddagger$ in Table~\ref{table:15bar_fits}, have spurious singularities if extrapolated too far below the constrained energy region. 
	For fits where such spurious poles appeared within $\sim1m_\pi$ below threshold, a penalty was added to push these well away in energy while ensuring the $\chi^2$ remained a smooth function of the parameters.
	The penalty function used was
	$$\chi_{\text{add}}^2(E_p)= \alpha \Bigg{(} \frac{f(E_p)}{1+e^{-f(E_p)} }\Bigg{)}^2 \, ,$$ 
	where $f(E_p) = \frac{E^2_p-E^2_{l}}{E^2_{h}-E^2_{l}}$, 
	$E_l = 0.42, E_h = 0.55$, $\alpha = 50$ and $E_p$ is the spurious pole mass, corresponding to a high penalty if the pole is near the physical scattering region and smoothly tending to zero as the pole location decreases to $E_l$.
	}
Despite the points appearing to lie roughly on a line, the \chisq\ are relatively poor
with fewer parameterisations yielding reasonable descriptions of the data (with $\chisq \leq 2$) then in the other flavour sectors.
As discussed in ~\mbox{\cite{Cheung:2020mql, Wilson:2023anv}}, this could be the result of strong correlations between energies levels being relatively more important in sectors with weak interactions;
imprecisely estimating the small eigenvalues of the correlation matrix can then lead to a disrupted fit.

\begin{table}
	\centering
	\resizebox{0.65\columnwidth}{!}{%
	\centering
			\begin{tabular}{ lc}
				\hline
			 	\multicolumn{1}{c}{\rule{0pt}{11.5pt} Parameterisation }  & \multicolumn{1}{c}{$\chi^2 / N_{\rm{dof}} \ $ } \\
				\hline
				  
				Effective range  																	    &                                    			 		  \\[5pt] 
				$\bm{(a)\ k} \textbf{ cot} \bm{\delta(k) = \frac{1}{a} + \frac{1}{2}rk^2}$            & $\bm{\frac{12.46}{9-2}= 1.78^{\ddag}} $                           \\[8pt]
				$\hphantom{----} \hookrightarrow$ {\footnotesize SVD cut $\kappa = 10^{-2}$}	        & $\frac{12.43}{9-2-1}= 2.07$           		  \\[8pt]
				$\hphantom{----} \hookrightarrow$ {\footnotesize SVD cut $\kappa = 5\times10^{-2}$}    	& $\frac{6.20}{9-2-3}= 1.55^{\ddag} $           		  \\[8pt]
				$\hphantom{----} \hookrightarrow$ {\footnotesize 	Uncorrelated}		                & $\frac{6.05}{9-2}= 0.86^{\ddag} $           		   	  \\[8pt]

				$(b)\ k \cot \delta(k) = \frac{1}{a} + \frac{1}{2}rk^2 + P_2k^4 $ 			            & \grey{$\frac{12.46}{9-3} = 2.08$}					      \\[8pt]

				\hline
				$K$-matrix with Chew-Mandelstam phase-space                             						           &                           			  \\[5pt] 
				$(c)\ K(s) = \gamma_0$                      								           & \grey{$\frac{17.85}{9-1} = 2.23$ }                       \\[8pt]
				$\bm{(d)\ K(s) = \gamma_0 + \gamma_1 s}$       			                       & $\bm{\frac{12.46}{9-2} = 1.78}$        		          \\[8pt]
				$\hphantom{----} \hookrightarrow$ {\footnotesize SVD cut $\kappa = 10^{-2}$}	       & $\frac{12.41}{9-2-1}= 2.07 $                	  \\[8pt]
				$\hphantom{----} \hookrightarrow$ {\footnotesize SVD cut $\kappa = 5\times 10^{-2}$}   & $\frac{4.85}{9-2-3}= 1.21 $           		   	          \\[8pt]
				$\hphantom{----} \hookrightarrow$ {\footnotesize 	Uncorrelated}		               & $\frac{5.73}{9-2}= 0.82 $           		   	          \\[8pt]
			
				$(e)\ K(s) = \gamma_0 + \gamma_1 s +\gamma_2 s^2$       				   & \grey{$\frac{12.46}{9-3} = 2.08$}                        \\[8pt]

				$(f)\ K^{-1}(s) = c_0 + c_1 s$                        	     		     	   & $\frac{12.45}{9-2} = 1.78$          			          \\[8pt]

				$(g)\ K^{-1}(s) = c_0 + c_1 s + c_2 s^2$                 				   & \grey{$\frac{12.45}{9-3} = 2.08$}  		              \\[8pt]

				\hline
				$K$-matrix with simple phase-space                       						           &                                    			      \\[5pt] 
				
				$(h)\ K(s) = \gamma_0 $                             			       				   & $\frac{14.87}{9-1} = 1.86$          			          \\[8pt]
	
				$(i)\ K(s) = \gamma_0 + \gamma_1 s$                         				       & $\frac{12.45}{9-2} = 1.78$          			          \\[8pt]

				$(j)\ K(s) = \gamma_0 + \gamma_1 s +\gamma_2 s^2$    				       & \grey{$\frac{12.71}{9-3} = 2.12$}        				  \\[8pt]

				$(k)\ K^{-1}(s) = c_0 + c_1 s $                        	  			     	   & $\frac{12.46}{9-2} = 1.78$           			          \\[8pt]

				$(l)\ K^{-1}(s) = c_0 + c_1 s + c_2 s^2$               				       & \grey{$\frac{12.45}{9-3} = 2.08$}       				  \\[8pt]

				\hline  
			\end{tabular}
		}
		\caption{As Table~\ref{table:3bar_fits} but for the flavour $\overline{\mathbf{15}}$ sector.
				For each of the reference parameterisations (in bold), in addition to the default fit, fits with two different SVD cutoffs are presented along with an uncorrelated fit.
				 Fits in grey are omitted from Fig.~\ref{fig:15bar_kcot_all_fits_and_SVD_uncorr_fits}. 
                 Fits marked with $^{\ddag}$ correspond to parameterisations that can have spurious singularities if extrapolated far below the energy region constrained as discussed in the text.
				 }\label{table:15bar_fits}
\end{table}

One way to deal with this, as done in \cite{Cheung:2020mql, Wilson:2023anv}, is to remove modes corresponding to the small\ poorly determined eigenvalues.
Using singular value decomposition (SVD),
modes with eigenvalue $\lambda_{i} < \kappa\lambda_{\text{max}}$ can be removed from consideration, for some cutoff $\kappa\in[0,1]$ and where $\lambda_{\text{max}}$ is the largest eigenvalue;
the number of degrees of freedom (dof) is reduced by one for each removed mode. 
To assess the impact of the poorly determined eigenvalues of the correlation matrix, we choose two different cutoffs, 
$\kappa = 10^{-2}$ and $5\times 10^{-2}$, and compare the reference parameterisations fits with and without these cutoffs. 
As a more extreme approach, we also consider uncorrelated fits where the off-diagonal elements of the correlation matrix are set to zero.
The results of these tests are shown in Table~\ref{table:15bar_fits} and the corresponding amplitudes are plotted in the lower panels of Fig.~\ref{fig:15bar_kcot_all_fits_and_SVD_uncorr_fits}.
For the effective range parameterisation $(a)$ the fit remained stable under the different fits, whilst for the $K-$matrix $(d)$ 
the fit curve had a small shift upwards when using a cutoff of $\kappa=5\times 10^{-2}$ or when performing an uncorrelated fit. 
Whilst the \chisq\ did improve using a cutoff $\kappa=5\times 10^{-2}$ or when ignoring correlations, the conclusion of weak interaction and no poles in the energy region constrained remained the same.

\section{Discussion} 
\label{section:disc}
\label{section:pole_structure}

This study with $SU(3)_f$ symmetry and $m_\pi \approx 700$ MeV finds
a bound state in the flavour $\bar{\mathbf{3}}$ sector,
 a virtual bound state in the flavour $\mathbf{6}$ sector and only weak repulsion in the flavour $\overline{\mathbf{15}}$ sector.
 The flavour $\bar{\mathbf{3}}$ bound state has a mass equal to the energy of the lowest finite-volume energy in the relevant irreps.
 This energy level has dominant overlap with $q\bar{q}$ operators  and the removal of meson-meson operators results in a tiny change in its energy.
 It appears that this bound state is consistent with being a conventional quark model state.

\begin{figure}
	\centering
	\includegraphics[width=0.8\linewidth]{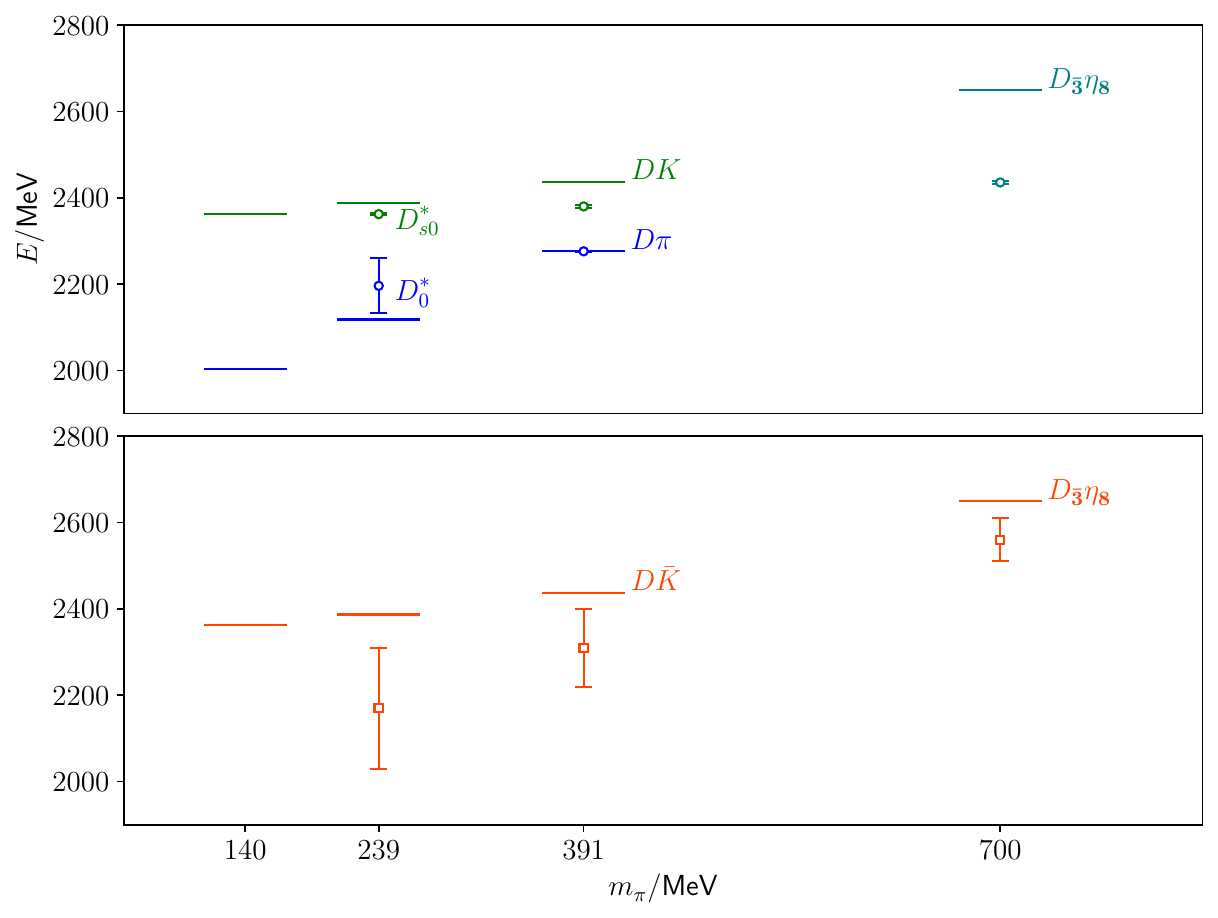}
	\caption{Pole locations as a function of pion mass from this work and Refs.~\cite{Moir:2016srx,Cheung:2020mql,Gayer:2021xzv} (points with error bars), and relevant thresholds (horizontal lines). 
			 Upper panel: $(S, I) = (0,\frac{1}{2})$ (blue) and $(1,0)$ (green) components of the flavour $\mathbf{\bar{3}}$ bound state.
			 Lower panel: $(-1,0)$ (orange) component of the flavour $\mathbf{6}$ virtual bound state.  
			 }\label{fig:runningpoles}
\end{figure}

Moving to smaller pion masses and breaking flavour symmetry, $SU(3)_f \rightarrow U(1)_S \times SU(2)_I$,
the evolution of the poles can be tracked
by looking at the (strangeness, isospin) $=(S,I)$ components of the $SU(3)_f$ irreps.
The flavour $\bar{\mathbf{3}}$ bound state splits into $(0, \frac{1}{2})$ and $(1, 0)$ components, 
identified as the $D^*_{0}$ and $D^*_{s0}$
in the charm-light and charm-strange sectors respectively.
The pole masses of the $D^*_{0}$ and $D^*_{s0}$ from previous Hadron Spectrum Collaboration lattice studies~\cite{Moir:2016srx, Gayer:2021xzv, Cheung:2020mql},
as well as the flavour  $\bar{\mathbf{3}}$ bound state found in this study,
are presented as a function of pion mass
in the upper panel of Fig.~\ref{fig:runningpoles}.
There it can be seen that the $D^*_{s0}$ remains a bound state below $DK$ threshold as the pion mass is lowered to its physical value
whilst the $D^*_{0}$ transitions to a resonance.
Naïvely following the trend of the lattice results in Fig.~\ref{fig:runningpoles} by eye might suggest that the $D^*_{0}$ mass at the physical pion limit is at odds with the current experimental value of 2343(10) MeV~\cite{ParticleDataGroup:2022pth}
from the Particle Data Group (PDG) review.
However, the PDG quotes the Breit-Wigner mass and it has been pointed out that this does not necessarily coincide with the pole mass for the $D^*_0$~\cite{vanBeveren:2003kd,vanBeveren:2006st,Gayer:2021xzv}, 
something that also explains why the quoted experimental mass of the $D^*_0$ is not lower than the mass of the $D^*_{s0}$, which is 2318.0(7) MeV~\cite{ParticleDataGroup:2022pth}.

The exotic flavour $\mathbf{6}$ virtual bound state at the $SU(3)_f$ point appears in three different flavour sectors when $SU(3)_f$ flavour symmetry is broken. Only its $(S,I) = (-1,0)$ component, corresponding to $D\bar{K}|_{I=0}$ scattering, has been reliably seen in previous lattice studies~\cite{Cheung:2020mql}. It appears as a virtual bound state for all three pion masses used and the pole mass is presented as a function of pion mass in the lower panel of Fig.~\ref{fig:runningpoles}.
Again, naïvely following the trend by eye to the physical pion mass suggests the $(S,I) = (-1,0)$ component of the $\mathbf{6}$ pole will remain a virtual bound state,  
consistent with the unitarised chiral perturbation theory (UChPT) prediction of a virtual bound state in the $D\bar{K}|_{I=0}$ channel at the physical pion mass~\cite{Albaladejo:2016lbb}. 
The finding of a flavour $\mathbf{6}$ pole in this study qualitatively agrees with the UChPT expectations of a near-threshold pole in the $SU(3)_f$ limit.
Ref.~\cite{Du:2017zvv} suggests that for sufficiently large light-quark masses there will be a bound state in this channel rather than a virtual bound state and our observation of a virtual bound state appears to be in tension with preliminary results from a more limited lattice study with $m_\pi \approx 600$ MeV where a bound state is claimed~\cite{Gregory:2021rgy}.

Another component of the $\mathbf{6}$ contributes to the $(0, \frac{1}{2})$ $D^*_0$ channel when flavour symmetry is broken. UChPT suggests that the $D^*_0$ has a two-pole structure with a lower pole corresponding to the $\mathbf{\bar{3}}$ in the flavour symmetry limit and the higher one corresponding to the $\mathbf{6}$~\cite{Kolomeitsev:2003ac, Albaladejo:2016lbb,Du:2020pui,Meissner:2020khl}.
Analysis within the UChPT framework~\cite{Du:2017zvv, Du:2019oki} of recent LHCb data \cite{LHCb:2016lxy, LHCb:2014ioa, LHCb:2015klp, LHCb:2015eqv, LHCb:2015tsv} gives supporting evidence for this two-pole structure.
While not excluded, it is less clear whether a two-pole structure is required to explain the lattice data at lower light-quark masses. 
HadSpec studies \cite{Moir:2016srx, Gayer:2021xzv} only robustly determine a single pole in the elastic-scattering region near the $D^*_0(2300)$, but a separate analysis
\cite{Asokan:2022usm} of the same lattice data suggested a second pole is required if $SU(3)_f$ flavour constraints are placed on the $K-$matrix.

The last sector where the $\mathbf{6}$ pole contributes when $SU(3)_f$ is broken is $(S,I)=(1,1)$, corresponding to  $D_s \pi- DK|_{I=1}$ scattering, and this has so far not been studied via lattice methods.
Studying it might give insight into the light-quark mass dependence of the $\mathbf{6}$ pole and how it manifests in a channel where it mixes with the weakly repulsive $\overline{\mathbf{15}}$.

\section{Summary and Outlook} 
\label{section:concl_and_outlook}

In summary, we have presented a lattice QCD study of elastic scattering with $J^P=0^+$ in the open-charm sector with $SU(3)_f$ flavour symmetry and $m_\pi \approx 700$ MeV. 
A bound state was found in the flavour $\bar{\mathbf{3}}$ sector with a pole mass in the range $m = 2432 - 2439$ MeV (roughly $210-220$ MeV below threshold), 
a virtual bound state was present in the flavour $\mathbf{6}$ sector with a pole location of $\sqrt{s_{\rm{pole}}} = 2510 -  2610$ MeV
(roughly $40-140$ MeV below threshold),
 and weak repulsive interactions were seen in the flavour $\overline{\mathbf{15}}$ sector.

Having determined the pole structure at the $SU(3)_f$ point from first principles, we have a more complete picture of the light-quark dependence of scattering in the scalar open-charm sector. 
Poles in different sectors at lower $m_\pi$ can be clearly seen as belonging to the same $SU(3)_f$ multiplets in the limit where up, down and strange quarks have equal mass (see Fig.~\ref{fig:runningpoles}).
The $D^*_{0}$ and the $D^*_{s0}$ correspond to the bound state seen in the flavour $\mathbf{\bar{3}}$, whilst a virtual bound state is seen in the exotic flavour $\mathbf{6}$.
Assuming $SU(3)_f$ flavour symmetry is not too badly broken at physical pion masses, 
we might expect to see this exotic $\mathbf{6}$ virtual bound state playing an important role in the charmed sector. 
With the (Strangeness, Isospin) $= (-1,0)$ component of the $\mathbf{6}$ pole reliably seen 
as a virtual bound state in $D\bar{K}|_{I=0}$ scattering at $m_{\pi} = 239, 391$ MeV \cite{Cheung:2020mql}
but still tension in identifying a two pole structure in the charmed anti-light $(0, \frac{1}{2})$ sector \cite{Moir:2016srx, Gayer:2021xzv, Asokan:2022usm}, 
we argue more work is needed to investigate how the components of this $\mathbf{6}$ pole manifests themselves with physical light-quark masses.

We suggest studies of $D_s \pi- DK|_{I=1}$ scattering at lower $m_\pi$ might reveal more information about how the $\mathbf{6}$ pole manifests itself towards the physical pion mass limit.
Furthermore, similar investigations of $DK/D\pi$ scattering at the $SU(3)_f$ flavour symmetry point can be done at different $m_\pi$ to probe the light-quark mass dependence of the dynamics in this sector.
Finally, exploration of the axial-vector $D^*\pi$ and $D^*K$ scattering channels in the open-charm, or analogous channels in the open-bottom sector, might reveal similar behaviour to that of the open-charm $J^P = 0^+$ sector and test the expected approximate heavy quark symmetry.

\acknowledgments
We thank our colleagues within the Hadron Spectrum Collaboration (\url{www.hadspec.org}), in particular Jozef Dudek, 
and also thank Christoph Hanhart and Feng-Kun Guo for useful discussions.
JDEY, CET and DJW acknowledge support from the U.K. Science and Technology Facilities Council (STFC) [grant numbers ST/T000694/1, ST/X000664/1].
DJW acknowledges support from a Royal Society University Research Fellowship.

The software codes {\tt Chroma}~\cite{Edwards:2004sx}, {\tt QUDA}~\cite{Clark:2009wm,Babich:2010mu} and {\tt Redstar}~\cite{Chen:2023zyy} were used.
Some software codes used in this project were developed with support from the U.S.\ Department of Energy, Office of Science, Office of Advanced Scientific Computing Research and Office of Nuclear Physics, Scientific Discovery through Advanced Computing (SciDAC) program; also acknowledged is support from the Exascale Computing Project (17-SC-20-SC), a collaborative effort of the U.S.\ Department of Energy Office of Science and the National Nuclear Security Administration.

This work used the Cambridge Service for Data Driven Discovery (CSD3), part of which is operated by the University of Cambridge Research Computing Service (\url{www.csd3.cam.ac.uk}) on behalf of the STFC DiRAC HPC Facility (\url{www.dirac.ac.uk}). The DiRAC component of CSD3 was funded by BEIS capital funding via STFC capital grants ST/P002307/1 and ST/R002452/1 and STFC operations grant ST/R00689X/1. Other components were provided by Dell EMC and Intel using Tier-2 funding from the Engineering and Physical Sciences Research Council (capital grant EP/P020259/1). This work also used the earlier DiRAC Data Analytic system at the University of Cambridge. This equipment was funded by BIS National E-infrastructure capital grant (ST/K001590/1), STFC capital grants ST/H008861/1 and ST/H00887X/1, and STFC DiRAC Operations grant ST/K00333X/1. DiRAC is part of the National E-Infrastructure.

Propagators used in this project were generated using DiRAC facilities, at Jefferson Lab under the USQCD Collaboration and the LQCD ARRA Project, under an ALCC award, using resources of the Oak Ridge Leadership Computing Facility at the Oak Ridge National Laboratory, which is supported by the Office of Science of the U.S. Department of Energy under Contract No. DE-AC05-00OR22725. and using resources of the National Energy Research Scientific Computing Center (NERSC), a DOE Office of Science User Facility supported by the Office of Science of the U.S. Department of Energy under Contract No. DE-AC02-05CH11231. The authors acknowledge the Texas Advanced Computing Center (TACC) at The University of Texas at Austin for providing HPC resources.
Gauge configurations were generated using resources awarded from the U.S. Department of Energy INCITE program at the Oak Ridge Leadership Computing Facility, the NERSC, the NSF Teragrid at the TACC and the Pittsburgh Supercomputer Center, using DiRAC facilities and at Jefferson Lab.

\vspace{0.5cm}

\noindent \textbf{Data Access Statement}
\vspace{0.2cm}

\noindent
The numerical values of the finite-volume energies along with statistical uncertainties and correlations, as well as full results of the fits, can be found in supplementary material.
Reasonable requests for other data can be directed to the authors and will be considered in accordance with the Hadron Spectrum Collaboration's policies on sharing data.

\appendix
\section{List of interpolating operators} \label{appendix:interpolating_list}

The operators used for each finite-volume irrep and volume are presented in this appendix.
Table~\ref{table:3bar_interpolating_list} shows the operators used for the flavour $\bar{\mathbf{3}}$ sector, Table~\ref{table:6f_rest_list} shows those used for at-rest irreps in the flavour $\mathbf{6}$ and $\overline{\mathbf{15}}$ sectors, and Table~\ref{table:6f_in_flight_list} shows those used for non-zero momentum irreps in the flavour $\mathbf{6}$ sector.
Meson-meson operators are labelled by meson pair and momentum type, i.e.\ $\mathbb{M}_{1[\vec{n}_1]}\mathbb{M}_{2[\vec{n}_2]}$,
and are ordered in the lists by corresponding non-interacting energy
\begin{align}
	a_tE_{\sf n.i.} = \sqrt{(a_tm_1)^2 + (a_tp_1)^2}+ \sqrt{(a_tm_2)^2 + (a_tp_2)^2} 
\end{align}
where $m_1$ and $m_2$ are the meson masses and 
$
(a_tp_i)^2 = \frac{1}{\xi^2}\Big{(}\frac{2\pi}{L/a_s}|\vec{n}_i|\Big{)}^2.
$

For the at-rest irreps all meson-meson operators up to the three-hadron threshold ($a_tE =  0.717$) are tabulated, 
whilst for $[100]A_1$ and $[110]A_1$ irreps all operators up to $a_tE_{\sf n.i.} = 0.70$ and $0.68$, respectively, are tabulated.
Any meson-meson operator not used in the basis when computing the correlation matrix is in grey in the tables. 
For each irrep, all meson-meson operators in the energy region of interest were included in the basis along with many extra to reduce contamination from higher energy states.
For the flavour $\bar{\mathbf{3}}$ sector all fermion-bilinears containing up to and including three gauge-covariant derivatives~\cite{Dudek:2010wm} were used,
and there are no fermion-bilinear operators for the $\mathbf{6}$ and $\overline{\mathbf{15}}$ sectors.

\begin{table}
		\centering
		\resizebox{0.5\columnwidth}{!}{%
		\begin{tabular}{ccc}
			$L/a_s=16$ & $L/a_s=20$ & $L/a_s=24$\\
			\hline
			$18\times \bar{\psi}\Gamma\psi$ & $18\times \bar{\psi}\Gamma\psi$ & $18\times \bar{\psi}\Gamma\psi$ \\
		    ${\D}_{[000]}{\etaoct}_{[000]}$ & ${\D}_{[000]}{\etaoct}_{[000]}$ & ${\D}_{[000]}{\etaoct}_{[000]}$ \\
			${\D}_{[100]}{\etaoct}_{[100]}$ & ${\D}_{[100]}{\etaoct}_{[100]}$ & ${\D}_{[100]}{\etaoct}_{[100]}$ \\
			${\D}_{[000]}{\etas}_{[000]}$ & ${\D}_{[000]}{\etas}_{[000]}$ & ${\D}_{[110]}{\etaoct}_{[110]}$ \\
			${\Dst}_{[000]}{\omoct}_{[000]}$ & ${\D}_{[110]}{\etaoct}_{[110]}$ & ${\D}_{[000]}{\etas}_{[000]}$ \\
			${\Dst}_{[000]}{\oms}_{[000]}$ & ${\D}_{[100]}{\etas}_{[100]}$ & ${\D}_{[111]}{\etaoct}_{[111]}$ \\
			${\D}_{[100]}{\etas}_{[100]}$ & ${\Dst}_{[000]}{\omoct}_{[000]}$ & ${\D}_{[100]}{\etas}_{[100]}$ \\
			${\D}_{[110]}{\etaoct}_{[110]}$ & ${\Dst}_{[000]}{\oms}_{[000]}$ & ${\D}_{[200]}{\etaoct}_{[200]}$ \\
			${\Dst}_{[100]}{\fzs}_{[100]}$ & ${\D}_{[111]}{\etaoct}_{[111]}$ & ${\Dst}_{[000]}{\omoct}_{[000]}$ \\
			${\Dst}_{[100]}{\omoct}_{[100]}\{2\}$ & ${\Dst}_{[100]}{\fzs}_{[100]}$ & ${\Dst}_{[000]}{\oms}_{[000]}$ \\
			${\Dst}_{[100]}{\oms}_{[100]}\{2\}$ & ${\D}_{[110]}{\etas}_{[110]}$ & ${\D}_{[110]}{\etas}_{[110]}$ \\
			${\D}_{[110]}{\etas}_{[110]}$ & ${\Dst}_{[100]}{\omoct}_{[100]}\{2\}$ & ${\Dst}_{[100]}{\fzs}_{[100]}$ \\
			${\D}_{[111]}{\etaoct}_{[111]}$ & ${\Dst}_{[100]}{\oms}_{[100]}\{2\}$ & ${\D}_{[210]}{\etaoct}_{[210]}$ \\
					& ${\D}_{[200]}{\etaoct}_{[200]}$ & ${\Dst}_{[100]}{\omoct}_{[100]}\{2\}$ \\
					& ${\Dst}_{[110]}{\fzs}_{[110]}$ & ${\Dst}_{[100]}{\oms}_{[100]}\{2\}$ \\
					& ${\D}_{[111]}{\etas}_{[111]}$ & ${\D}_{[111]}{\etas}_{[111]}$ \\
					& ${\Dst}_{[110]}{\omoct}_{[110]}\{3\}$ & ${\Dst}_{[110]}{\fzs}_{[110]}$ \\
					& ${\Dst}_{[110]}{\oms}_{[110]}\{3\}$ & ${\D}_{[211]}{\etaoct}_{[211]}$ \\
					& \grey{${\D}_{[210]}{\etaoct}_{[210]}$} & ${\Dst}_{[110]}{\omoct}_{[110]}\{3\}$ \\
					&         & ${\Dst}_{[210]}{\etaoct}_{[210]}$ \\
					&         & ${\Dst}_{[110]}{\oms}_{[110]}\{3\}$ \\
					&         & ${\D}_{[200]}{\etas}_{[200]}$ \\
					&         & ${\Dst}_{[111]}{\fzs}_{[111]}$ \\
					&         & ${\Dst}_{[111]}{\omoct}_{[111]}\{2\}$ \\
					&         & ${\Dst}_{[111]}{\oms}_{[111]}\{2\}$ \\
					&         & ${\D}_{[210]}{\etas}_{[210]}$ \\
					&         & ${\Dst}_{[211]}{\etaoct}_{[211]}$ \\
					&         & ${\Dst}_{[200]}{\fzs}_{[200]}$ \\
			\end{tabular}
		}
		\caption{Number of fermion-bilinear operators containing up to three gauge-covariant derivatives and list of meson-meson operators with corresponding non-interacting energies up to three-hadron threshold for the $[000]A_1^+$ irrep in the flavour $\mathbf{\bar{3}}$ sector on the $(L/a_s)^3 = 16^3$, $20^3$ and $24^3$ volumes.
                Multiplicities are labelled in $\{ \}$ whenever more than one.
		 		Operators not included in the basis when computing correlation functions are in grey.
                }
		\label{table:3bar_interpolating_list} 
\end{table}

\begin{table}
	\centering
	\resizebox{0.48\columnwidth}{!}{%
		\begin{tabular}[t]{ccc}
			\multicolumn{3}{c}{$[000]A_1^+$} \\
			\hline
			$L/a_s=16$ & $L/a_s=20$ & $L/a_s=24$\\
			\hline
			${\D}_{[000]}{\etaoct}_{[000]}$ & ${\D}_{[000]}{\etaoct}_{[000]}$ & ${\D}_{[000]}{\etaoct}_{[000]}$ \\
			${\D}_{[100]}{\etaoct}_{[100]}$ & ${\D}_{[100]}{\etaoct}_{[100]}$ & ${\D}_{[100]}{\etaoct}_{[100]}$ \\
			${\Dst}_{[000]}{\omoct}_{[000]}$ & ${\D}_{[110]}{\etaoct}_{[110]}$ & ${\D}_{[110]}{\etaoct}_{[110]}$ \\
			${\D}_{[110]}{\etaoct}_{[110]}$ & ${\Dst}_{[000]}{\omoct}_{[000]}$ & ${\D}_{[111]}{\etaoct}_{[111]}$ \\
			${\Dst}_{[100]}{\omoct}_{[100]}\{2\}$ & ${\D}_{[111]}{\etaoct}_{[111]}$ & ${\D}_{[200]}{\etaoct}_{[200]}$ \\
			${\D}_{[111]}{\etaoct}_{[111]}$ & ${\Dst}_{[100]}{\omoct}_{[100]}\{2\}$ & ${\Dst}_{[000]}{\omoct}_{[000]}$ \\
					& ${\D}_{[200]}{\etaoct}_{[200]}$ & ${\D}_{[210]}{\etaoct}_{[210]}$ \\
					& ${\Dst}_{[110]}{\omoct}_{[110]}\{3\}$ & ${\Dst}_{[100]}{\omoct}_{[100]}\{2\}$ \\
					& \grey{${\D}_{[210]}{\etaoct}_{[210]}$} & ${\D}_{[211]}{\etaoct}_{[211]}$ \\
					&         & ${\Dst}_{[110]}{\omoct}_{[110]}\{3\}$ \\
					&         & ${\Dst}_{[210]}{\etaoct}_{[210]}$ \\
					&         & ${\Dst}_{[111]}{\omoct}_{[111]}\{2\}$ \\
					&         & ${\Dst}_{[211]}{\etaoct}_{[211]}$ \\
			\end{tabular}
	}
	\resizebox{0.48\columnwidth}{!}{%
		\begin{tabular}[t]{ccc}
			\multicolumn{3}{c}{$[000]T_1^-$} \\
			\hline
			$L/a_s=16$ & $L/a_s=20$ & $L/a_s=24$\\
			\hline
			${\D}_{[100]}{\etaoct}_{[100]}$ & ${\D}_{[100]}{\etaoct}_{[100]}$ & ${\D}_{[100]}{\etaoct}_{[100]}$ \\
			${\Dst}_{[100]}{\etaoct}_{[100]}$ & ${\Dst}_{[100]}{\etaoct}_{[100]}$ & ${\Dst}_{[100]}{\etaoct}_{[100]}$ \\
			${\D}_{[110]}{\etaoct}_{[110]}$ & ${\D}_{[110]}{\etaoct}_{[110]}$ & ${\D}_{[110]}{\etaoct}_{[110]}$ \\
			${\D}_{[100]}{\omoct}_{[100]}$ & ${\Dst}_{[110]}{\etaoct}_{[110]}\{2\}$ & ${\D}_{[111]}{\etaoct}_{[111]}$ \\
			${\Dst}_{[110]}{\etaoct}_{[110]}\{2\}$ & ${\D}_{[111]}{\etaoct}_{[111]}$ & ${\Dst}_{[110]}{\etaoct}_{[110]}\{2\}$ \\
			${\Dst}_{[100]}{\omoct}_{[100]}\{4\}$ & ${\D}_{[100]}{\omoct}_{[100]}$ & ${\D}_{[100]}{\omoct}_{[100]}$ \\
			${\D}_{[111]}{\etaoct}_{[111]}$ & ${\Dst}_{[111]}{\etaoct}_{[111]}$ & ${\D}_{[200]}{\etaoct}_{[200]}$ \\
			${\D}_{[110]}{\omoct}_{[110]}\{2\}$ & ${\Dst}_{[100]}{\omoct}_{[100]}\{4\}$ & ${\Dst}_{[111]}{\etaoct}_{[111]}$ \\
					& ${\D}_{[110]}{\omoct}_{[110]}\{2\}$ & ${\D}_{[110]}{\omoct}_{[110]}\{2\}$ \\
					& ${\D}_{[200]}{\etaoct}_{[200]}$ & ${\D}_{[210]}{\etaoct}_{[210]}\{2\}$ \\
					& ${\Dst}_{[110]}{\omoct}_{[110]}\{7\}$ & ${\Dst}_{[100]}{\omoct}_{[100]}\{4\}$ \\
					& ${\Dst}_{[200]}{\etaoct}_{[200]}$ & ${\Dst}_{[200]}{\etaoct}_{[200]}$ \\
					& ${\D}_{[111]}{\omoct}_{[111]}$ & ${\D}_{[111]}{\omoct}_{[111]}$ \\
					& \grey{${\D}_{[210]}{\etaoct}_{[210]}\{2\}$} & ${\D}_{[211]}{\etaoct}_{[211]}\{2\}$ \\
					&         & ${\Dst}_{[110]}{\omoct}_{[110]}\{7\}$ \\
					&         & ${\Dst}_{[210]}{\etaoct}_{[210]}\{4\}$ \\
					&         & ${\D}_{[200]}{\omoct}_{[200]}$ \\
					&         & ${\Dst}_{[111]}{\omoct}_{[111]}\{5\}$ \\
					&         & ${\Dst}_{[211]}{\etaoct}_{[211]}\{2\grey{, 2}\}$ \\
			\end{tabular}
	}
	\centering
	\resizebox{0.48\columnwidth}{!}{%
		\begin{tabular}[t]{ccc}
			\multicolumn{3}{c}{$[000]E^+$} \\
			\hline
			$L/a_s=16$ & $L/a_s=20$ & $L/a_s=24$\\
			\hline
			${\D}_{[100]}{\etaoct}_{[100]}$ 			& ${\D}_{[100]}{\etaoct}_{[100]}$ 				& ${\D}_{[100]}{\etaoct}_{[100]}$ \\
			${\Dst}_{[000]}{\omoct}_{[000]}$ 			& ${\D}_{[110]}{\etaoct}_{[110]}$ 				& ${\D}_{[110]}{\etaoct}_{[110]}$ \\
			${\D}_{[110]}{\etaoct}_{[110]}$ 			& ${\Dst}_{[110]}{\etaoct}_{[110]}$ 			& ${\Dst}_{[110]}{\etaoct}_{[110]}$ \\
			${\Dst}_{[110]}{\etaoct}_{[110]}$ 			& ${\Dst}_{[000]}{\omoct}_{[000]}$ 				& ${\D}_{[200]}{\etaoct}_{[200]}$ \\
			${\Dst}_{[100]}{\omoct}_{[100]}\{3\}$ 		& ${\Dst}_{[111]}{\etaoct}_{[111]}$ 			& ${\Dst}_{[000]}{\omoct}_{[000]}$ \\
			${\D}_{[110]}{\omoct}_{[110]}$ 				& ${\Dst}_{[100]}{\omoct}_{[100]}\{3\}$ 		& ${\Dst}_{[111]}{\etaoct}_{[111]}$ \\
														& ${\D}_{[110]}{\omoct}_{[110]}$ 				& ${\D}_{[110]}{\omoct}_{[110]}$ \\
														& ${\D}_{[200]}{\etaoct}_{[200]}$ 				& ${\D}_{[210]}{\etaoct}_{[210]}\{2\}$ \\
														& ${\Dst}_{[110]}{\omoct}_{[110]}\{5\}$ 		& ${\Dst}_{[100]}{\omoct}_{[100]}\{3\}$ \\
														& ${\D}_{[111]}{\omoct}_{[111]}$ 				& ${\D}_{[111]}{\omoct}_{[111]}$ \\
														&  \grey{${\D}_{[210]}{\etaoct}_{[210]}\{2\}$} 	& ${\D}_{[211]}{\etaoct}_{[211]}$ \\
														&        										& ${\Dst}_{[110]}{\omoct}_{[110]}\{5\}$ \\
														&       										& ${\Dst}_{[210]}{\etaoct}_{[210]}\{2\}$ \\
														&         										& ${\Dst}_{[111]}{\omoct}_{[111]}\{3\}$ \\
														&         										& ${\Dst}_{[211]}{\etaoct}_{[211]}\{1, \grey{2}\}$ \\
														&       									    &  	\\
			\end{tabular}
	}
	\resizebox{0.48\columnwidth}{!}{%
		\begin{tabular}[t]{ccc}
			\multicolumn{3}{c}{$[000]T_2^+$} \\
			\hline
			$L/a_s=16$ & $L/a_s=20$ & $L/a_s=24$\\
			\hline
			${\Dst}_{[100]}{\etaoct}_{[100]}$			    & ${\Dst}_{[100]}{\etaoct}_{[100]}$ 			& ${\Dst}_{[100]}{\etaoct}_{[100]}$ \\
			${\Dst}_{[000]}{\omoct}_{[000]}$ 				& ${\D}_{[110]}{\etaoct}_{[110]}$ 				& ${\D}_{[110]}{\etaoct}_{[110]}$ \\
			${\D}_{[110]}{\etaoct}_{[110]}$ 				& ${\Dst}_{[110]}{\etaoct}_{[110]}\{2\}$ 		& ${\D}_{[111]}{\etaoct}_{[111]}$ \\
			${\D}_{[100]}{\omoct}_{[100]}$ 					& ${\Dst}_{[000]}{\omoct}_{[000]}$ 				& ${\Dst}_{[110]}{\etaoct}_{[110]}\{2\}$ \\
			${\Dst}_{[110]}{\etaoct}_{[110]}\{2\}$ 			& ${\D}_{[111]}{\etaoct}_{[111]}$ 				& ${\D}_{[100]}{\omoct}_{[100]}$ \\
			${\Dst}_{[100]}{\omoct}_{[100]}\{3\}$ 			& ${\D}_{[100]}{\omoct}_{[100]}$ 				& ${\Dst}_{[000]}{\omoct}_{[000]}$ \\
			${\D}_{[111]}{\etaoct}_{[111]}$ 				& ${\Dst}_{[111]}{\etaoct}_{[111]}$ 			& ${\Dst}_{[111]}{\etaoct}_{[111]}$ \\
			${\D}_{[110]}{\omoct}_{[110]}\{2\}$ 			& ${\Dst}_{[100]}{\omoct}_{[100]}\{3\}$ 		& ${\D}_{[110]}{\omoct}_{[110]}\{2\}$ \\
															& ${\D}_{[110]}{\omoct}_{[110]}\{2\}$   		& ${\D}_{[210]}{\etaoct}_{[210]}$ \\
															& ${\Dst}_{[110]}{\omoct}_{[110]}\{7\}$ 		& ${\Dst}_{[100]}{\omoct}_{[100]}\{3\}$ \\
															& ${\Dst}_{[200]}{\etaoct}_{[200]}$ 			& ${\Dst}_{[200]}{\etaoct}_{[200]}$ \\
															& ${\D}_{[111]}{\omoct}_{[111]}$ 				& ${\D}_{[111]}{\omoct}_{[111]}$ \\
															& \grey{${\D}_{[210]}{\etaoct}_{[210]}$}		& ${\D}_{[211]}{\etaoct}_{[211]}\{2\}$ \\
															&         										& ${\Dst}_{[110]}{\omoct}_{[110]}\{7\}$ \\
															&        									    & ${\Dst}_{[210]}{\etaoct}_{[210]}\{5\}$ \\
															&      										    & ${\D}_{[200]}{\omoct}_{[200]}$ \\
															&         										& ${\Dst}_{[111]}{\omoct}_{[111]}\{5\}$ \\
															&        									    & ${\Dst}_{[211]}{\etaoct}_{[211]}\{2,\grey{2}\}$ \\
															&        									    &  	\\
			\end{tabular}
	}
	\caption{As Table \ref{table:3bar_interpolating_list} but for $[000]\Lambda^P$ irreps in the flavour $\mathbf{6}$ and $\mathbf{\overline{15}}$ sectors.
			 Degeneracies split in the form $\{n,\grey{m}\}$ indicate how many operators of the degeneracy were included (n) and not included (m)
			 when computing the matrix of correlation functions.}
		\label{table:6f_rest_list} 
	\end{table}

	\begin{table}
		\centering
		\resizebox{0.8\columnwidth}{!}{%
		\begin{tabular}{ccc|ccc}
			\multicolumn{3}{c|}{$[100]A_1$ } & \multicolumn{3}{c}{$[110]A_1$ } \\
			\hline
			\multicolumn{3}{c|}{$L/a_s=24$} & \multicolumn{3}{c}{$L/a_s=24$} \\
			\hline
			${\D}_{[100]}{\etaoct}_{[000]}$ & ${\Dst}_{[210]}{\etaoct}_{[110]}$ & ${\D}_{[211]}{\etaoct}_{[210]}$ & ${\D}_{[110]}{\etaoct}_{[000]}$ & ${\D}_{[110]}{\etaoct}_{[200]}$ & ${\Dst}_{[210]}{\etaoct}_{[111]}\{3\}$ \\
			${\D}_{[000]}{\etaoct}_{[100]}$ & ${\D}_{[211]}{\etaoct}_{[111]}$ & ${\Dst}_{[110]}{\omoct}_{[100]}\{5\}$ & ${\D}_{[100]}{\etaoct}_{[100]}$ & ${\Dst}_{[100]}{\etaoct}_{[111]}$ & ${\Dst}_{[110]}{\omoct}_{[000]}\{3\}$ \\
			${\D}_{[110]}{\etaoct}_{[100]}$ & ${\D}_{[110]}{\etaoct}_{[210]}$ & ${\Dst}_{[210]}{\etaoct}_{[200]}$ & ${\D}_{[000]}{\etaoct}_{[110]}$ & ${\D}_{[210]}{\etaoct}_{[111]}$ & ${\Dst}_{[100]}{\etaoct}_{[210]}$ \\
			${\D}_{[100]}{\etaoct}_{[110]}$ & ${\D}_{[110]}{\omoct}_{[100]}$ & ${\D}_{[110]}{\omoct}_{[111]}$ & ${\D}_{[111]}{\etaoct}_{[100]}$ & ${\Dst}_{[200]}{\etaoct}_{[110]}$ & ${\D}_{[110]}{\omoct}_{[110]}\{3\}$ \\
			${\D}_{[200]}{\etaoct}_{[100]}$ & ${\D}_{[210]}{\etaoct}_{[200]}$ & \grey{${\D}_{[300]}{\etaoct}_{[200]}$} & ${\Dst}_{[100]}{\etaoct}_{[100]}$ & ${\D}_{[100]}{\etaoct}_{[210]}$ &\grey{${\D}_{[221]}{\etaoct}_{[111]}$} \\
			${\Dst}_{[110]}{\etaoct}_{[100]}$ & ${\Dst}_{[100]}{\omoct}_{[000]}\{2\}$ & ${\D}_{[210]}{\etaoct}_{[211]}$ & ${\D}_{[110]}{\etaoct}_{[110]}$ & \grey{${\D}_{[220]}{\etaoct}_{[110]}$} & ${\D}_{[210]}{\etaoct}_{[210]}$ \\
			${\D}_{[111]}{\etaoct}_{[110]}$ & ${\D}_{[100]}{\omoct}_{[110]}$ & ${\Dst}_{[100]}{\omoct}_{[110]}\{5\}$ & ${\D}_{[210]}{\etaoct}_{[100]}$ & ${\D}_{[100]}{\omoct}_{[100]}$ & ${\Dst}_{[100]}{\omoct}_{[100]}\{5\}$ \\
			${\D}_{[110]}{\etaoct}_{[111]}$ & ${\D}_{[200]}{\etaoct}_{[210]}$ & ${\Dst}_{[200]}{\etaoct}_{[210]}$ & ${\D}_{[100]}{\etaoct}_{[111]}$ & \grey{${\Dst}_{[211]}{\etaoct}_{[110]}\{3\}$} & ${\D}_{[100]}{\omoct}_{[111]}$ \\
			${\Dst}_{[100]}{\etaoct}_{[110]}$ & ${\Dst}_{[000]}{\omoct}_{[100]}\{2\}$ & ${\D}_{[210]}{\omoct}_{[110]}$ & ${\Dst}_{[111]}{\etaoct}_{[100]}$ & ${\D}_{[111]}{\etaoct}_{[210]}$ & ${\D}_{[210]}{\omoct}_{[100]}$ \\
			${\D}_{[210]}{\etaoct}_{[110]}$ & ${\D}_{[111]}{\etaoct}_{[211]}$ & \grey{${\D}_{[220]}{\etaoct}_{[210]}$} & ${\D}_{[200]}{\etaoct}_{[110]}$ & ${\Dst}_{[110]}{\etaoct}_{[200]}$ & ${\D}_{[200]}{\etaoct}_{[211]}$ \\
			${\D}_{[100]}{\etaoct}_{[200]}$ & \grey{${\Dst}_{[211]}{\etaoct}_{[111]}$} & ${\Dst}_{[200]}{\omoct}_{[100]}\{2\}$ & ${\Dst}_{[110]}{\etaoct}_{[110]}\{3\}$ & ${\D}_{[111]}{\omoct}_{[100]}$ & ${\Dst}_{[000]}{\omoct}_{[110]}\{3\}$ \\
			${\Dst}_{[111]}{\etaoct}_{[110]}$ & ${\Dst}_{[110]}{\etaoct}_{[210]}$ & ${\Dst}_{[111]}{\etaoct}_{[211]}$ & ${\Dst}_{[210]}{\etaoct}_{[100]}$ & ${\D}_{[211]}{\etaoct}_{[200]}$ & ${\Dst}_{[111]}{\etaoct}_{[210]}\{3\}$ \\
			${\Dst}_{[110]}{\etaoct}_{[111]}$ & ${\D}_{[111]}{\omoct}_{[110]}$ & ${\Dst}_{[111]}{\omoct}_{[110]}\{5\}$ & ${\D}_{[211]}{\etaoct}_{[110]}$ & ${\D}_{[110]}{\etaoct}_{[211]}$ & ${\D}_{[200]}{\omoct}_{[110]}$ \\
					&         &         &        &         & ${\Dst}_{[111]}{\omoct}_{[100]}\{2\grey{, 3}\}$ \\
			\end{tabular}
		}
		\caption{As Table \ref{table:6f_rest_list} but for $[100]A_1$ and $[110]A_1$ irreps on the $24^3$ volume in the flavour $\mathbf{6}$ sector. 
				 }
		\label{table:6f_in_flight_list} 
	\end{table}

\clearpage

\bibliographystyle{JHEP-2}
\bibliography{refs.bib}

\end{document}